\documentclass[aip,jcp,amsmath,amssymb,reprint,numeric]{revtex4-1}
\usepackage{graphicx}% Include figure files
\usepackage{subfigure}
\usepackage{dcolumn}% Align table columns on decimal point
\usepackage{bm}% bold math
\usepackage{mathrsfs}
\usepackage[dvipsnames]{xcolor}
\usepackage{xcolor}
\usepackage[normalem]{ulem}
\usepackage{caption}
\begin{document}

\title{Dynamic signature of the thermodynamic transition in a novel mean field system\\} 
\author{Ehtesham Anwar}
\address{\textit{Polymer Science and Engineering Division, CSIR-National Chemical Laboratory, Pune-411008, India}}
\affiliation{\textit{Academy of Scientific and Innovative Research (AcSIR), Ghaziabad 201002, India}}
\author{Ujjwal Kumar Nandi}
\address{\textit{Polymer Science and Engineering Division, CSIR-National Chemical Laboratory, Pune-411008, India}}
\affiliation{\textit{Academy of Scientific and Innovative Research (AcSIR), Ghaziabad 201002, India}}
\affiliation{\textit{Department of Physics, Kyoto University, Kyoto 606-8502, Japan}}
\author{Palak Patel}
\address{\textit{Polymer Science and Engineering Division, CSIR-National Chemical Laboratory, Pune-411008, India}}
\affiliation{\textit{Academy of Scientific and Innovative Research (AcSIR), Ghaziabad 201002, India}}
%\address{\textit{Department of Chemical, Materials and Production Engineering, University of Naples Federico II, Piazzale Tecchio 80, Napoli 80125, Italy}}
\author{Sanket Kumawat}
\address{\textit{Polymer Science and Engineering Division, CSIR-National Chemical Laboratory, Pune-411008, India}}
\affiliation{\textit{Academy of Scientific and Innovative Research (AcSIR), Ghaziabad 201002, India}}
\author{Sarika Maitra Bhattacharyya}
\email{mb.sarika.nclp@nic.in}
\address{\textit{Polymer Science and Engineering Division, CSIR-National Chemical Laboratory, Pune-411008, India}}
\affiliation{\textit{Academy of Scientific and Innovative Research (AcSIR), Ghaziabad 201002, India}}

%\date{December 2020}

\begin{abstract}
 Understanding the connection between thermodynamics and dynamics in glass-forming liquids remains a central challenge in condensed matter physics. In this study, we investigate a novel model system that enables a continuous crossover from a standard three dimensional liquid to a fully connected mean field like system by introducing pseudo neighbours. These pseudo neighbours enhance the effective connectivity of the system without altering its local structure. While their presence slows down the dynamics, they influence thermodynamic properties even more significantly. In particular, the configurational entropy obtained via thermodynamic integration vanishes at a temperature much higher than the temperature where the dynamics begin to slow down, leading to a clear breakdown of the Adam Gibbs relation. To uncover a possible dynamical signature of this thermodynamic transition, we analyse bond breakage dynamics. Unlike real-real bonds, which decay similarly in both the parent Kob Andersen model and its mean field variant, real-pseudo bonds exhibit long lived, persistent behaviour with strong temperature dependence. These bonds do not fully decay over time, leading to a finite saturation value of the bond breakage correlation function. Remarkably, we show that the number of surviving pseudo bonds can be analytically estimated and correlates directly with the thermodynamic transition temperature $T_{K}$. We propose a phenomenological relation between $T_K$ and the number of surviving pseudo-bonds, establishing a novel link between thermodynamic and dynamic observables. Our results suggest that these persistent pseudo bonds serve as a robust dynamical signature of the thermodynamic transition, and the system might have properties analogous to those of randomly bonded ultrastable glasses.

\end{abstract}

\maketitle
\section{Introduction}
The dramatic and non Arrhenius increase in viscosity and relaxation times observed in supercooled liquids as they are cooled toward the glass transition is one of the most prominent yet least understood phenomena in condensed matter physics \cite{Debenedetti_book,Debenedetti2001,Angell1995,Berthier2011,sastry_inherent_1998}. As the temperature decreases, these liquids fall out of equilibrium and become glassy, but the precise mechanism driving this kinetic slowdown remains elusive. At the core of this puzzle lies a fundamental question: Is the glass transition purely kinetic in nature, governed by the dynamics of defects or local rearrangements \cite{Chandler2010}, or does it signify an underlying thermodynamic phase transition characterized by a vanishing configurational entropy at a finite temperature?\cite{Adam1965,wolynes_lubchenko}

Several theoretical frameworks have been developed to describe aspects of this slowing down. Mode Coupling Theory (MCT) provides a mean field approach that captures many features of the dynamics at moderately supercooled temperatures, predicting a power law divergence of relaxation times at a critical temperature, $T_c$\cite{Goetze2009,Kob1995,Das2004}. However, both simulations and experiments have revealed that, rather than diverging, the dynamics undergo a smooth crossover below $T_c$, indicating the involvement of activated processes not accounted for by MCT\cite{sarika_pnas,sastry_inherent_1998,Nandi_2020,chong}. The activated dynamics was described by the Random First-Order Transition (RFOT) theory\cite{ Kirkpatrick1989, Lubchenko2007,wolynes_lubchenko, Cavagna2009}, which suggested that the dramatic slowing down is associated with the vanishing of configurational entropy $S_c$ at the Kauzmann temperature, $T_K$, beyond which the number of accessible amorphous states becomes negligible.  The Adam Gibbs (AG) relation, which connects the divergence of the relaxation time to the vanishing of the configurational entropy, offers a phenomenological basis for this connection\cite{Adam1965,Richert1998}.

Despite the conceptual appeal of RFOT and AG, direct verification remains difficult due to the practical challenges of equilibrating glass forming systems at low temperatures. Usually, the thermodynamic transition temperature is calculated by extrapolating the configurational entropy values to lower temperatures. However, recent studies have introduced alternate approaches, such as randomly pinning a fraction of particles to raise $T_K$ to accessible regimes and probe the thermodynamic nature of the glass transition in equilibrium conditions \cite{Biroli_phase_diagram,walter_original_pinning,smarajit_chandan_dasgupta_original_pinning,parisi_jamming_pinned_system,ozawa_prl_pinned,palak_pinned}. These investigations have shown promise in mapping out the transition in the temperature and pinning fraction phase space. Interestingly, in many studies where the $T_{K}$ is obtained from extrapolation, the AG relationship was found to hold \cite{atryee_prl_2014,atreyee_2016,manoj_2015,srikant_adam_2013,shiladitya_prl_2012,Anshul_prl_2017,Sastry2001}. However, for the pinned system, it was found that the correlation between the self overlap dynamics and the configurational entropy does not follow the AG relationship \cite{walter_original_pinning,ujjwal_thermodynamics}.
The studies showed that the collective overlap function shows the signature of the thermodynamic transition.\cite{walter_original_pinning,reply_by_kob} 

In a recent study, a similar breakdown of the AG relationship was observed \cite{ujjwal_thermodynamics}. That work, involving some of us, introduced a model system enabling a controlled and continuous crossover from a standard three dimensional liquid to a fully connected, mean field like system. This transition was achieved by increasing the number of interacting neighbours per particle through the introduction of pseudo neighbours, thereby enhancing the system’s effective connectivity. Although similar in spirit to the Mari–Kurchan (MK) model\cite{mari_kurchan}, a key distinction lies in structural integrity. While the MK model results in a gas-like radial distribution function (rdf) at high mean field coupling, the present model always preserves the local structure \cite{ujjwal_mean_field}.

It was shown that adding more pseudo neighbours slows down the dynamics \cite{ujjwal_mean_field}. However, pseudo neighbours have a much stronger influence on thermodynamics \cite{ujjwal_mean_field}. Specifically, configurational entropy calculated via thermodynamic integration (TI) was found to vanish at a significantly higher temperature than that at which the dynamics showed considerable slowing down, leading to a clear breakdown of the AG relation. Unlike in pinned particle systems, where the collective dynamics reflected the thermodynamic transition\cite{walter_original_pinning,reply_by_kob}, in this mean field model, neither the self nor the collective overlap functions exhibited any signature of the transition \cite{ujjwal_thermodynamics}.

Motivated by these observations, the present work explores whether dynamical signatures of the thermodynamic transition can still be recovered in other ways. In particular, earlier results indicated that entropy calculated via the phenomenological two-phase thermodynamic (2PT) method\cite{2pt_method1,2pt_method2,2pt_method_3}, based on the velocity auto-correlation function (VACF), remains higher than that obtained via the TI method and continues to obey the AG relation. Interestingly, we find that pseudo neighbours have a moderate effect on the VACF, highlighting that local dynamics is weakly perturbed by the pseudo neighbours. However, pseudo neighbours have a stronger effect on long time behaviour, notably delaying the onset of diffusion.

This observation points to the persistence of bonds between real and pseudo neighbours. While these long lived connections do not influence local relaxation dynamics, they significantly affect the long time diffusive regime. In this article, we present a detailed analysis of bond breakage dynamics in both the mean field system and its parent Kob Andersen (KA) model. We find that, unlike the real–real bonds (in both models), which decay over time, real–pseudo bonds exhibit a long time plateau, indicating the presence of a finite number of persistent bonds. We show that this number can be statistically estimated from the structure of the liquid.

A key outcome of this study is the derivation of a phenomenological expression connecting the Kauzmann temperature, $T_{K}$ to the number of persistent final pseudo bonds. This result suggests that a dynamic signature of the thermodynamic transition is encoded in the saturation value of the bond breakage correlation function, offering a novel perspective on the coupling between structure, dynamics, and thermodynamics in these glassy systems.

The rest of the article is organised in the following way. The simulation details are in section \ref{sec_details}. In section \ref{result}, we present the result, while in section \ref{conclusion}, we summarise and conclude. This paper contains two Appendix sections at the end.

\section{System and simulation details}
\label{sec_details}

The mean field system consists of $N$ particles that interact through a typical short-range potential. Additionally, each particle also interacts with $k$ “pseudo neighbours”—particles that may not be spatially close. Therefore, the overall interaction potential in the system includes both local and nonlocal contributions, given by,

\begin{eqnarray}
U_{\rm tot}(r_{1},..r_{N})&=&\sum_{i=1}^{N}\sum_{j>i}^{N}u(r_{ij})+\frac{1}{2}\sum_{i=1}^{N}\sum_{j=1}^{k}u^{\rm pseudo}(r_{ij}) \nonumber \; \;
\label{eq1}\\
&=&U+U^{\rm pseudo}_{k} \qquad .
\label{e}
\end{eqnarray}
\noindent

The first term on the right-hand side represents the standard Lennard-Jones (LJ) interaction between particles. The second term accounts for the interactions between each particle and its pseudo neighbours. We consider a binary Lennard-Jones (LJ) system with $80\%$ type A and $20\%$ type B particles. The pairwise interaction is

\begin{equation}
u(r_{ij})=4\epsilon_{ij}\Big[\Big(\frac{\sigma_{ij}}{r_{ij}}\Big)^{12}-\Big(\frac{\sigma_{ij}}{r_{ij}}\Big)^6\Big] \quad,
\end{equation}
\noindent
where $r_{ij}$ is the distance between the particles, $\sigma_{ij}$ is the effective diameter of the particle and $\epsilon_{ij}$ is the interaction strength. We use $\sigma_{AA}$ and $\epsilon_{AA}$ as length and energy units, setting the Boltzmann constant $k_B=1$. Parameter values (from Ref.~\citenum{Kob1995}) are:  $\sigma_{AA}=1.0$, $\sigma_{AB}=0.8$,  $\sigma_{BB}$=0.88, $\epsilon_{AA}$=1.0 $\epsilon_{AB}$=1.5, and $\epsilon_{BB}$=0.5, making the system a good glass former. The potential is truncated and shifted at $r_c=2.5\sigma_{ij}$, with particle mass $m_A=m_B=1$, and time units defined by $\sqrt{m_A \sigma^2_{AA}/\epsilon_{AA}}$.

These pseudo neighbour interactions are described using a modified, shifted, and truncated version of the Lennard-Jones potential,\cite{ujjwal_mean_field}

\begin{eqnarray}
u^{\rm pseudo}(r_{ij})&=&u(r_{ij}-L_{ij}) \nonumber \\
&=&4\epsilon_{ij}\Big[\Big(\frac{\sigma_{ij}}{r_{ij}-L_{ij}}\Big)^{12}-\Big(\frac{\sigma_{ij}}{r_{ij}-L_{ij}}\Big)^6\Big] \quad 
\label{ka_model}
\end{eqnarray}

In our simulations, we enforce that any two particles interact either through $u(r_{ij})$ or through $u^{\rm pseudo}(r_{ij})$ but not both. This rule guides the selection of pseudo neighbours and the corresponding $L_{ij}$ values for a given configuration equilibrated under the potential $u$. For each particle $i$, we randomly select $k$ distinct particles $j$ such that $L_{box}/2>r_{ij}>r_c$ where $L_{box}/2$ is the half box length. We then assign a $L_{ij}$ value and accept the pseudo neighbours if $(r_{ij}-L_{ij})\ge 1$.
In the earlier studies $L_{ij}$ was taken as a variable with a distribution \cite{ujjwal_mean_field,ujjwal_thermodynamics}. In the present study, to understand the effect of the $L_{ij}$, we keep it fixed for all pairs, $L_{ij}=L$. Importantly, once a particle $j$ is assigned as a pseudo neighbour of $i$, the relationship is mutual, so $i$ is also considered a pseudo neighbour of $j$. The system defined in this way is compatible with standard simulation methods.
We perform NVT molecular dynamics (MD) simulations\cite{dfrenkel} in a cubic box using the velocity rescaling method, with $N=2744$ particles at a density $\rho=1.2$. The time step for integration is set to $\Delta t=0.005$. We study systems with $k=0,4,6,12,20,24$, and $28$ at two different values of $L$, $L=2.5$ and $L=3$.

To analyse the effect of system size on the behaviour of the mean field system, we conduct two complementary sets of simulations. In one set while keeping $k=6$ fixed, N is varied thus changing $\rho_{k}=\frac{k}{V-\frac{4}{3}\pi L^{3}}$: $N=1411$ at  $\rho_k=0.0054$, $N=2744$ at $\rho_k=0.0026$, $N=4967$ at $\rho_k=0.0015$, and $N=9261$ at $\rho_k=0.0008$. In the second set, 
both N and $k$ are varied in a way that $\rho_{k}=0.0054$ remains constant: $N=1411$ and $k=6$, $N=2744$ and $k=12$, $N=4967$ and $k=22$, and $N=15634$ and $k=70$. These simulations enable a systematic study of finite size effects, both at fixed and variable pseudo interaction densities.

\section{Results}
 \label{result}

\begin{figure}[h!]
  \centering
 \includegraphics[width=0.5\textwidth]{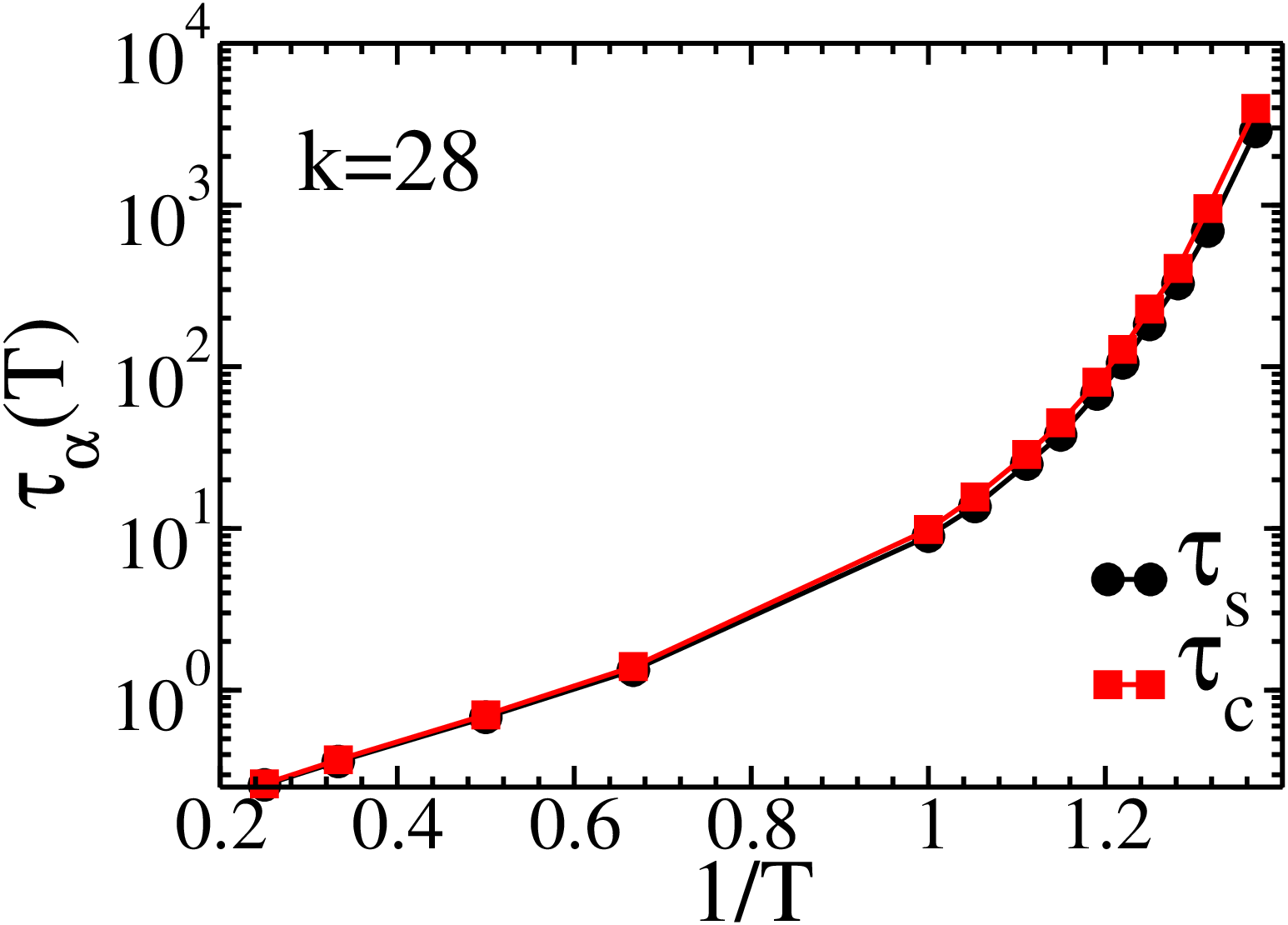}
 \caption{Temperature dependence of the $\alpha$ relaxation timescale of the self and collective overlap functions, $\tau_{s}$ and $\tau_{c}$ respectively, for $k=28$ and $L=2.5$ system. Both self and collective dynamics follow each other.} 
 \label{Fig-overlap}
 \end{figure}

As discussed in the Introduction, a previous study involving some of us showed that when the configurational entropy of the mean field system is computed using the thermodynamic integration (TI) method, it vanishes at a temperature significantly higher than the temperature at which the dynamics slow down appreciably \cite{ujjwal_thermodynamics}. This behaviour closely resembles that observed in systems with randomly pinned particles \cite{walter_original_pinning}. However, in pinned systems, the collective overlap function exhibits a clear signature of the thermodynamic transition \cite{walter_original_pinning}, whereas in the mean field system, both the self and collective overlap functions continue to decay, indicating that the system remains dynamically active even as the configurational entropy vanishes \cite{ujjwal_thermodynamics}. In Fig.\ref{Fig-overlap}, we plot the alpha relaxation times, $\tau_{s}$ obtained from the self overlap function $Q_{s}(t)$, and $\tau_{c}$ obtained from the collective overlap function $Q_{c}(t)$ (defined in the Appendix \ref{overlap}). Unlike the behaviour seen in soft and hard pinned systems \cite{ehtesham_2024,walter_original_pinning}, where the self and collective dynamics decouple, we observe that in the mean field model, they closely track each other.

To understand the AG breakdown, in the earlier study, the entropy was also calculated via the Two-Phase Thermodynamics (2PT) method, a heuristic but reliable and widely used technique \cite{ujjwal_thermodynamics}. It predicted a $T_{K}$ value lower than that obtained via the TI method \cite{ujjwal_thermodynamics}.  It was also observed that as $k$ increases, the discrepancy between the entropy values obtained from the TI and 2PT methods becomes more pronounced. Interestingly, the Adam Gibbs relationship was found valid when the entropy was calculated using the 2PT method \cite{ujjwal_thermodynamics}.

\begin{figure}[h!]
 \centering
 \includegraphics[width=0.5\textwidth]{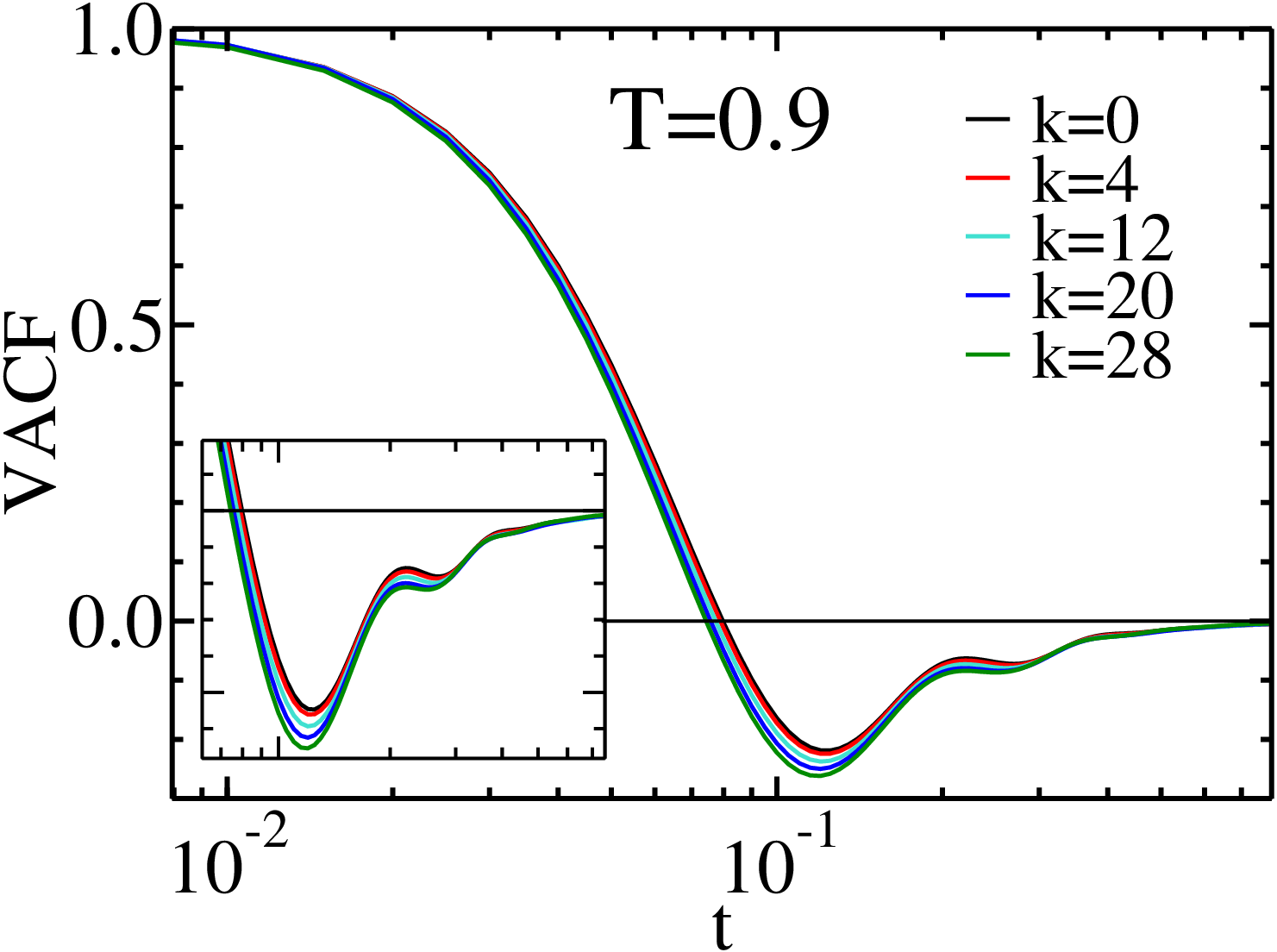}
 \caption{The normalised velocity autocorrelation function (VACF) as a function of time for $L=2.5$ and $k=0,4,12,20$, and $28$ systems at $T=0.9$. The time at which the VACF first decays to zero shows a weak {$k$} dependence.}
 \label{velo_corr}
 \end{figure}

\begin{figure}[h!]
 \centering
 \includegraphics[width=0.5\textwidth]{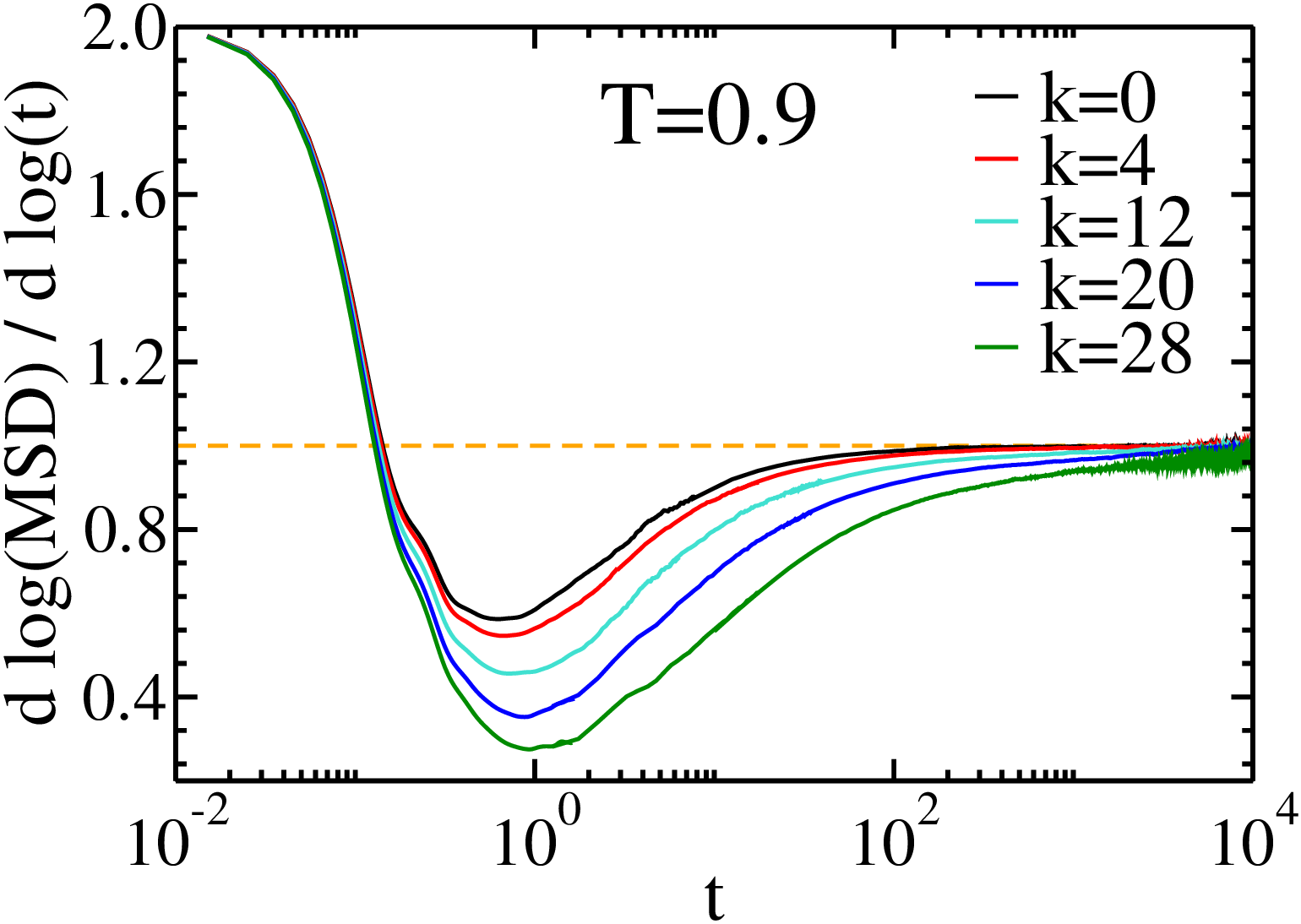}
 \caption{The time derivative of the logarithm of the mean square displacement (MSD) as a function of time for $L=2.5$ and $k=0,4,12,20$ and $28$ systems at $T=0.9$. When the value of the derivative reaches unity, the system enters a diffusive regime. The onset of the diffusive regime shows a strong $k$ dependence.} 
 \label{msd_der}
 \end{figure}

In the 2PT method, the entropy is calculated from the dynamics, namely the velocity autocorrelation function (VACF) given by $C_v(t) = \big <v_i(t).v_i(0)\big > =1/N\sum_{i=1}^{N} v_i(t).v_i(0)$. 
In Fig.\ref{velo_corr}, we plot the normalised VACF, and we show that with an increase in $k$, the VACF has a deeper minimum, which is commensurate with the increased caging and slowing down of the dynamics. However, for the different systems, the time over which the velocity autocorrelation function decays to zero weakly depends on the number of pseudo neighbours. Thus, the entropy calculated from the (VACF) also has a weaker $k$ dependence. This is similar to the weaker $k$ dependence of the overlap function and the validity of the AG relationship when entropy is calculated using the 2PT method. On the other hand as reported earlier \cite{ujjwal_thermodynamics} we observe that the time over which the mean square displacement, $MSD=R^2(t)=\frac{1}{N} \sum_{i} \langle({\bf{r}}_i(t)-{\bf{r}}_i(0))^2 \rangle$,  becomes diffusive, obtained from the time $\frac{dlog(MSD)}{d log(t)} =1$ has a much stronger $k$ dependence (Fig.\ref{msd_der}). For large values of $k$, the system does not become diffusive at low temperatures within the simulation timescale. Note that in the mean field system, particles are caged not only by their real neighbours, as in conventional liquids, but also by the pseudo neighbours, which introduce long range, persistent constraints. The cage formed by pseudo neighbours is significantly larger in spatial extent than that formed by real neighbours. As a result, a particle may undergo physical displacement once the real neighbour's cage relaxes, leading to the decay of the self and collective overlap function, and also the velocity auto-correlation function (VACF). However, the broader cage formed by the pseudo neighbours may continue to restrict motion on longer time scales, especially in terms of the onset of diffusion.

\begin{figure}[h!]
 \centering
 \includegraphics[width=0.5\textwidth]{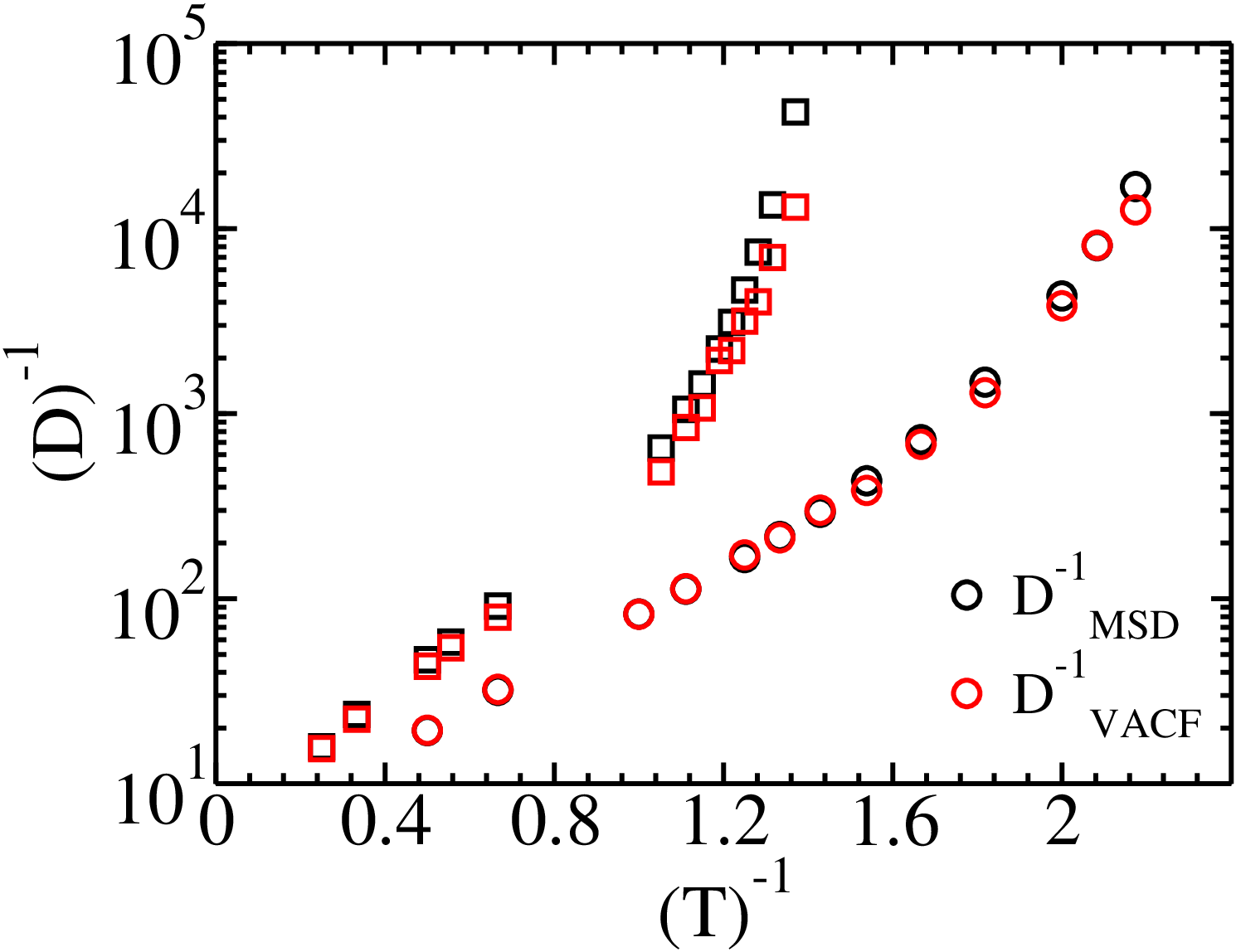}
 \caption{Temperature dependence of the diffusion coefficient, $D_{VACF}$, obtained from the velocity autocorrelation function and $D_{MSD}$, obtained from the mean square displacement for $k=28$ and $L=2.5$ system (square) and KA ($k=0$) system (circle). In the KA system, both the diffusion values are similar; however, in the $k=28$ system,  $D_{MSD}$ is smaller than $D_{VACF}$, and with the decrease in temperature, this difference increases.} 
 \label{diff_frm_vacf_msd}
 \end{figure}

This observation raises an important question: Do the pseudo neighbours merely delay the onset of diffusion, or do they also influence the diffusion constant? To address this, we compute and compare the diffusion constant\cite{Hansen_and_McDonald} 
$D$ using two independent approaches: (i) from the long-time limit of the mean squared displacement (MSD), $D_{MSD}= \lim_{t\rightarrow \infty} \frac{R^2(t)}{6t}$ where the expression is valid only in the diffusive regime of the MSD, which sets in at the time $\frac{d log(MSD)}{d log(t)} =1$ and (ii) from the Green–Kubo relation involving the integral of the VACF, $ D_{VACF}= 1/3 \int_{t=0}^{t=\infty} <v_i(0).v_i(t)>dt$, where $t=\infty$ is the time the VACF has decayed to zero.
In Fig.\ref{diff_frm_vacf_msd}, we compare the diffusion constants and find that for the Kob–Andersen (KA) model, the diffusion values obtained from both methods agree well across the temperature range studied. However, in the mean field system, we observe a clear discrepancy: the $D_{MSD}$ is consistently lower than $D_{VACF}$, and this difference grows at lower temperatures. This indicates that the pseudo neighbours have a stronger influence on the MSD, which captures long time displacements, than on the VACF, which probes local dynamics. To understand this effect more systematically, one might study cage breakage dynamics, {\it i.e}., how the cage formed by surrounding particles dissolves over time. However, defining cage breakage is nontrivial, as it depends on specifying how many cage forming neighbours must move to declare the cage "broken," introducing inherent ambiguity. To avoid this ambiguity, we instead focus on bond breakage dynamics, which offers a more precise measure of structural relaxation. Moreover, it has been shown in a previous work that below the onset temperature, the timescale associated with the system’s transition to diffusive behaviour closely tracks the timescale extracted from bond breakage correlations \cite{smarjit_timescale}. Thus, studying bond breakage provides a robust way to understand the role of pseudo neighbours in constraining long time dynamics.

\subsection{Bond breakage time correlation function}
The bond breakage (BB) time correlation function is defined in the following
way\cite{Ryoichi_Yamamoto1997,smarjit_timescale}. At $t = t_0$ a
pair of particle $i$ and $j$ is considered to be bonded if
\begin{equation}
r_{ij}(t_0) = |\vec{r}_i(t_0) - \vec{r}_j(t_0)| \le 1.4\sigma_{11}.
\label{BB_RR}
\end{equation}

If $r_{ij}(t+t_{0}) \le 1.4\sigma_{11}$ (the position of the first minima in $g_{11}(r)$) the bond
is said to have survived at time $t$. We calculate the remaining number of initial bonds at time $t$, and the bond breakage time correlation function is defined
as the ratio of this number to the initial number of bonds.

\begin{figure}[h!]
 \centering
 \includegraphics[width=0.5\textwidth]{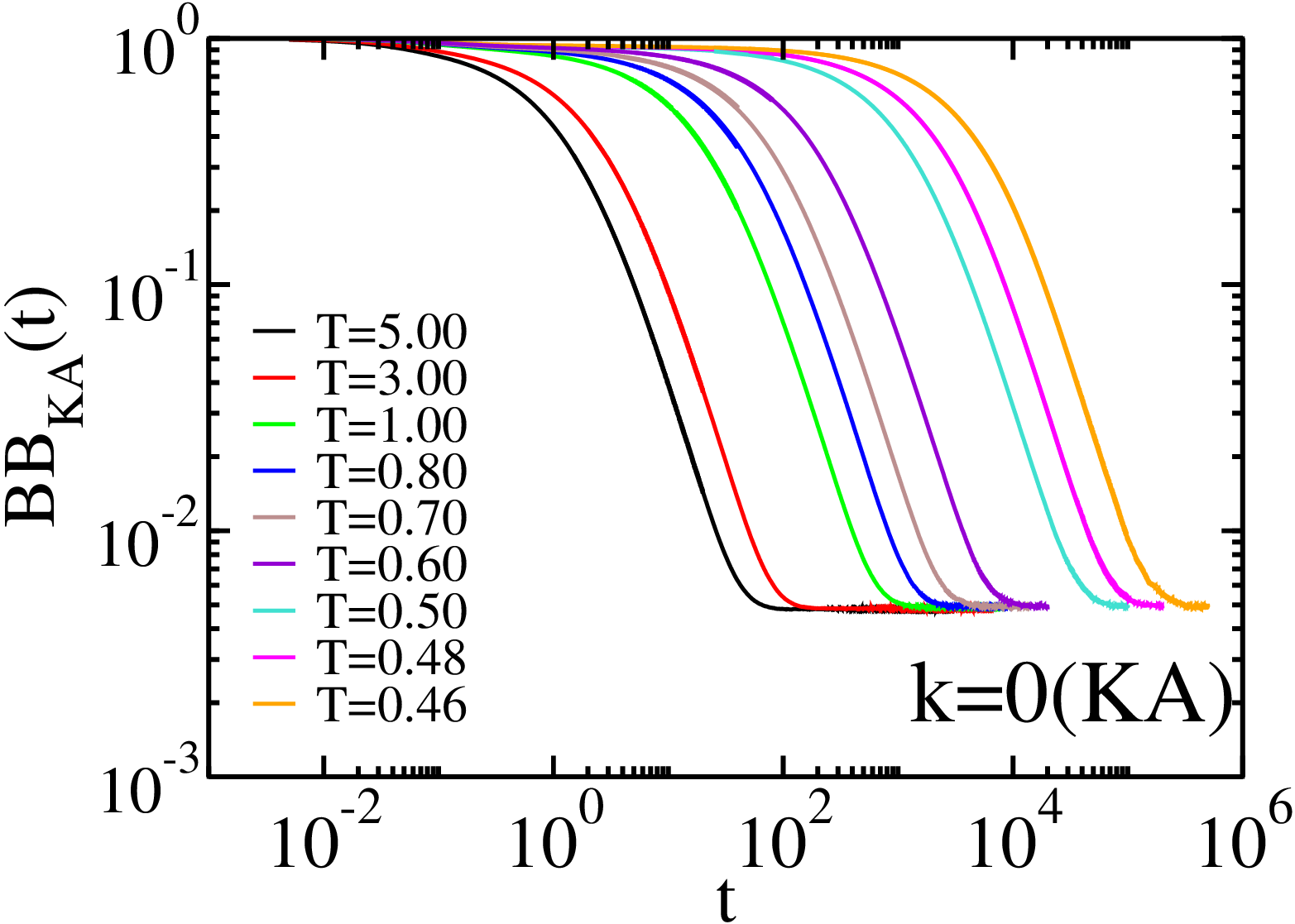}
 \caption{$BB_{KA}(t)$ the bond breakage time correlation function for KA ($k=0$) system at different temperatures. The Bond breakage correlation function at a long time does not go to zero but saturates at a finite value, and this saturation value is almost temperature independent.} 
 \label{bond_break_KA}
 \end{figure}

As shown in Fig.\ref{bond_break_KA}, the bond breakage time correlation function for a KA model, $BB_{KA}(t)$ initially decays, but instead of going to zero, it saturates at a certain value, $S_{KA}=BB_{KA}(t \rightarrow \infty)$. This value, where it saturates, is almost temperature independent. We will later discuss the physical significance of this saturation value. 

While working with the mean field model, we have bifurcated the bond breakage time correlation function into two parts: the real-real bond breakage ($BB_{RR}$)  correlation function which is defined by Eq.\ref{BB_RR} and the real-pseudo bond breakage ($BB_{RP}$) correlation function. 

The real-pseudo bond breakage correlation function is defined in the following way.
At $t = t_0$ a pair of particle $i$  and $k$  is considered to be bonded if
\begin{equation}
r'_{ik}(t_0) = |\vec{r}_i(t_0) - \vec{r}_k(t_0)| - L \le 2.5\sigma_{11}.
\end{equation}

\begin{figure}[h!]
 \centering
 \includegraphics[width=0.5\textwidth]{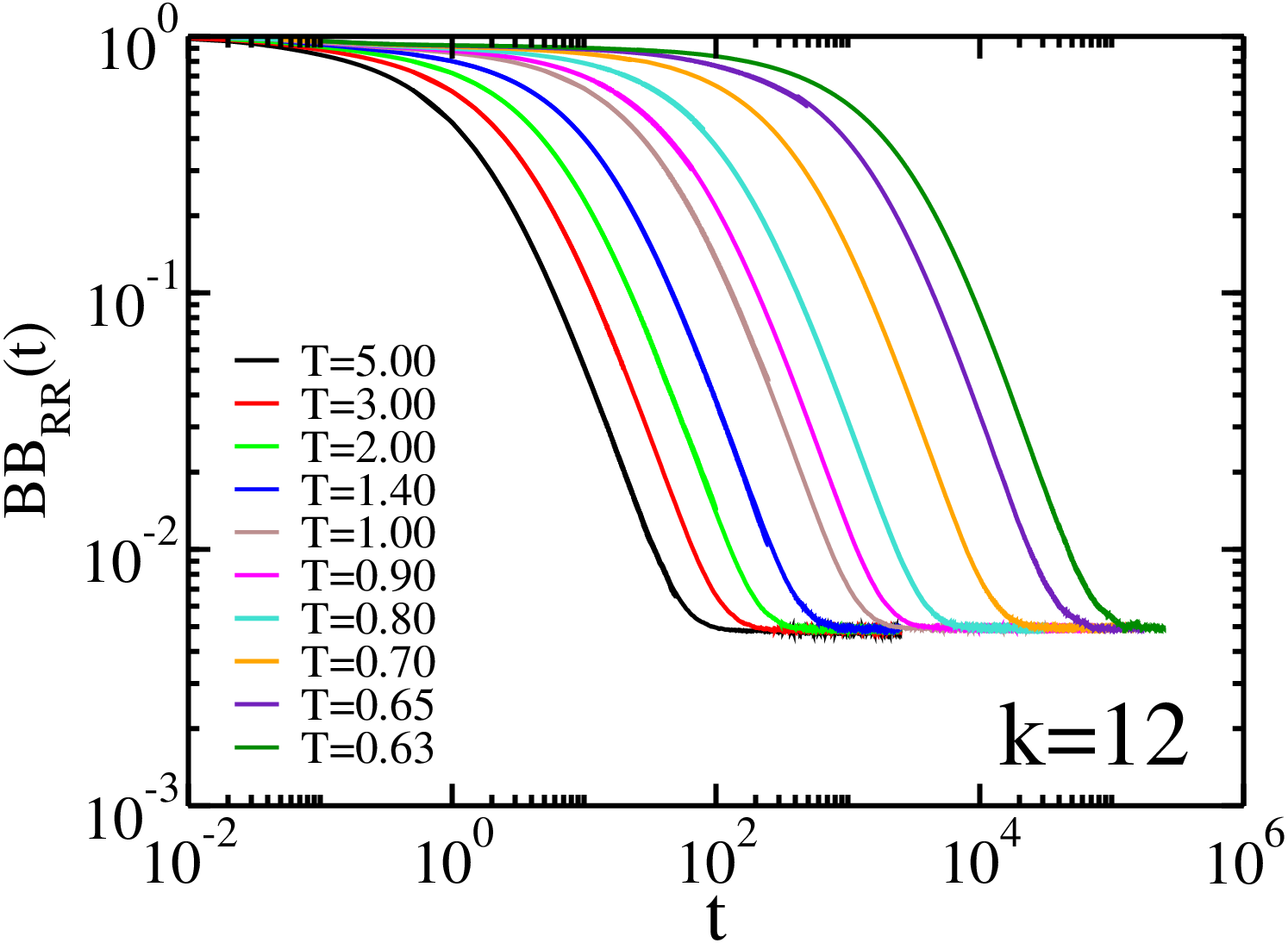}
 \caption{$BB_{RR}(t)$, real-real bond breakage time correlation function for $k=12$ and $L=2.5$ system at different temperatures. The saturation value is almost temperature independent.} 
 \label{real_real_bb}
 \end{figure}

\begin{figure}[h!]
 \centering
 \includegraphics[width=0.5\textwidth]{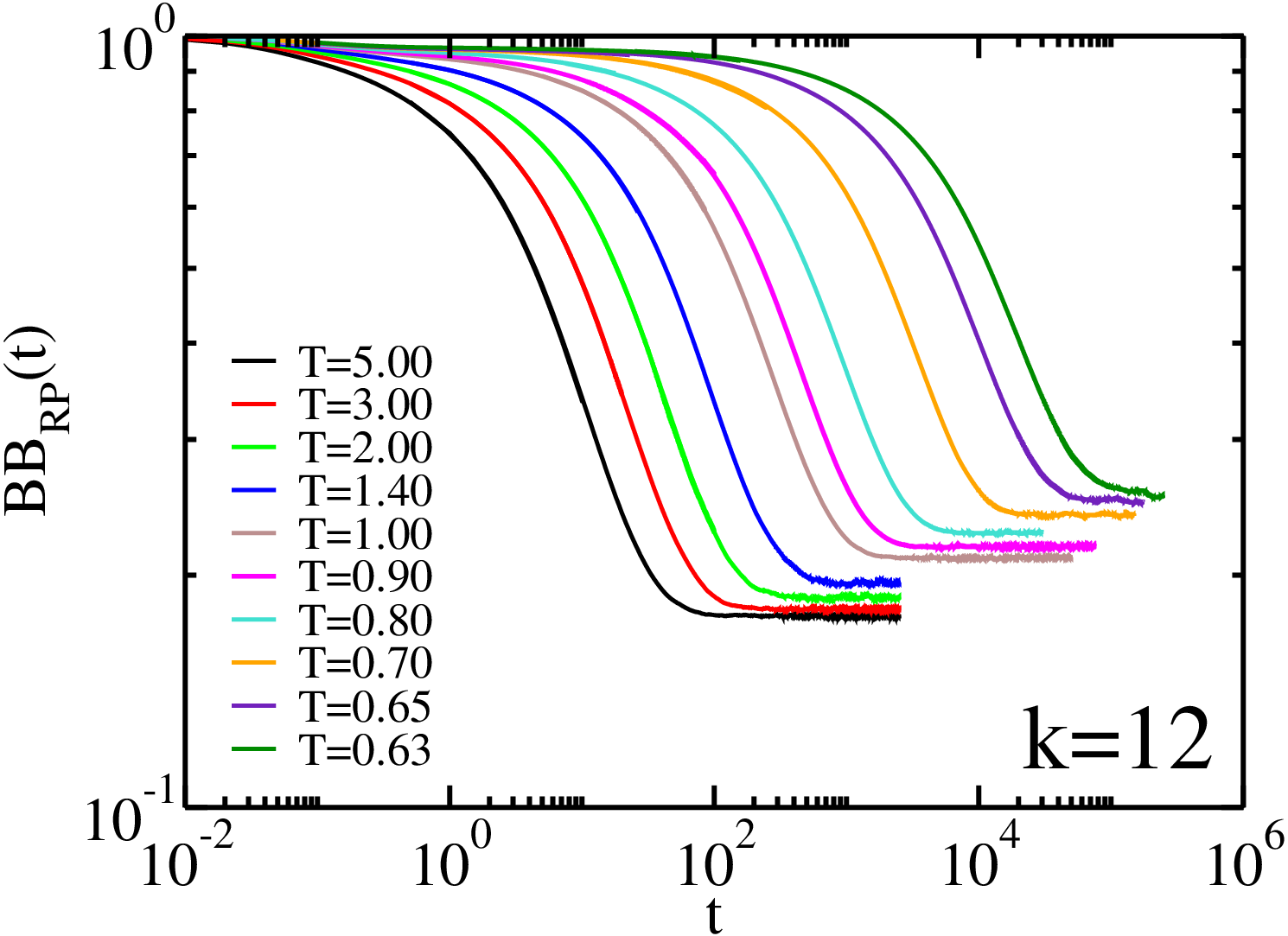}
 \caption{$BB_{RP}(t)$, real-pseudo bond breakage time correlation function for $k=12$ and $L=2.5$ system at different temperatures. The saturation value of the real-pseudo bond breakage correlation has a strong temperature dependence.} 
 \label{real_pseudo_bb}
 \end{figure}

\begin{figure}[h!]
 \centering
 \includegraphics[width=0.5\textwidth]{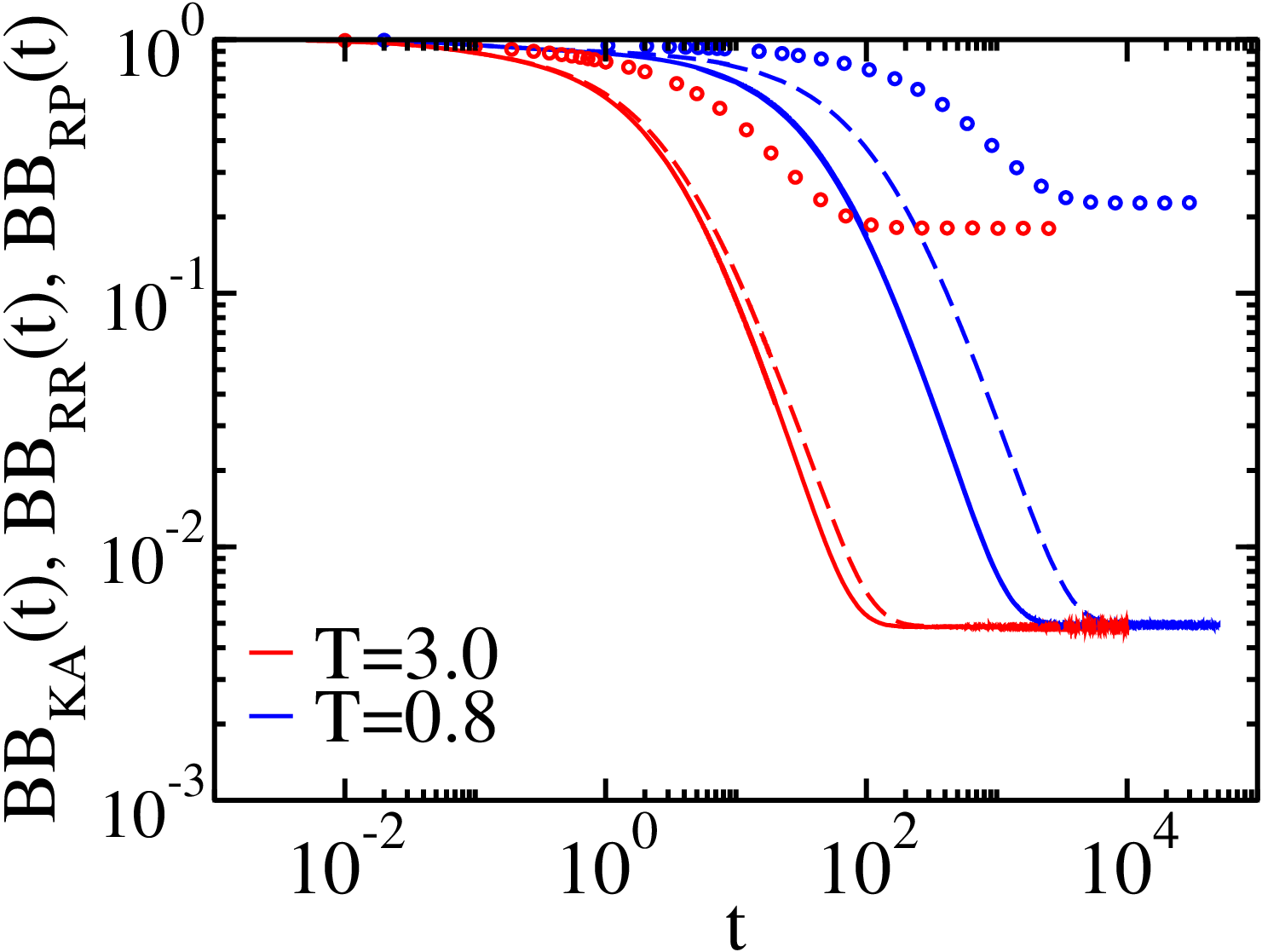}
 \caption{Comparison between different bond breakage time correlation functions. $BB_{KA}(t)$, bond breakage time correlation function, for $KA$ (k=0) system (solid lines), $BB_{RR}(t)$, real-real bond breakage time correlation function for mean field system  (dashed lines)  and $BB_{RP}(t)$, real-pseudo bond breakage time correlation function (symbols)  for mean field system at $T=3.0$ and $T=0.8$. The time evolution of $BB_{KA}(t)$ and $BB_{RR}(t)$ is different; $BB_{RR}(t)$ is slower, but the saturation values are the same. When compared to $BB_{KA}(t)$ and $BB_{RR}(t)$, we find $BB_{RP}(t)$ have an even slower decay and saturates at a higher value.} 
 \label{comp_BB}
 \end{figure}

If $r^{'}_{ik}(t+t_{0}) \le 2.5\sigma_{11}$, the bond is said to have survived at time $t$. Note that since the pseudo rdf is quite similar to $exp(-\beta u(r))$ and has no first peak and minima \cite{ujjwal_mean_field}, we take the full range of the interaction. We have checked that the nature of the result remains the same when we consider other cutoff values of $r^{'}$ to describe the survival of the bond. We calculate the number of initial bonds that
survive at time $t$, and the bond breakage time correlation function is defined
as the ratio of this number to the initial number of bonds.
 In Fig.\ref{real_real_bb} and Fig.\ref{real_pseudo_bb} we plot the time evolution of the bond breakage time correlation function of the real-real and the real-pseudo bonds, respectively, at different temperatures. In Fig.\ref{comp_BB} we also show a comparative behaviour of the time evolution of the bonds in the KA model, the real-real bonds and the real-pseudo bonds in the mean field system, both at high and low temperatures. 
We find that compared to $BB_{KA}(t)$,$BB_{RR}(t)$ has a slower decay, and the difference in the timescale of decay increases with a decrease in temperature (Fig.\ref {comp_BB}). This is because the pseudo neighbours present in the system slow down the overall dynamics. A similar effect was observed in the overlap function \cite{ujjwal_mean_field}. Interestingly although the dynamics is slower the saturation value of the $BB_{RR}(t)$, $S_{RR}=BB_{RR}(t\rightarrow \infty)$ is same as that for the pure KA model {\it i.e.} $S_{RR}=S_{KA}$. It has a small value and is almost temperature independent. This implies that the number of surviving real bonds in the mean field system is the same as the number of surviving bonds in a KA system. Compared to both $BB_{KA}(t)$ and $BB_{RR}(t)$ the $BB_{RP}(t)$  shows an even slower decay (Fig. \ref{comp_BB}).  Again, at low temperatures, the difference in the timescale of decay becomes wider. This clearly shows that in the mean field system, the pseudo bond breaking is slower than the real bond breaking. What is most important is that the real-pseudo bond breakage time correlation function's saturation value, $S_{RP}=BB_{RP}(t\rightarrow \infty)$, is significant and strongly dependent on temperature, as shown in Fig.\ref{real_pseudo_bb} and Fig.\ref{comp_BB}. This implies that compared to the real bonds, the surviving pseudo bonds are much more significant in number and increase with a decrease in temperature. 

\subsection{Statistical analysis of the initial and surviving bonds}
In order to understand what leads to the surviving bonds in the system, we do the following analysis.
As discussed in an earlier work involving some of us \cite{ujjwal_mean_field}, the probability of finding a pseudo particle within a distance of $2.5\sigma$ of a real particle is given by  
\begin{equation}
P_{RP}=\frac{\int_{V_{\rm acc}} d{\bf r} \, e^{-\beta u(r-L)}y(r)}{\int_{V_{\rm acc}} d{\bf r}\,e^{-\beta u(r-L)}}.
\label{prob_RP}
\end{equation}
\noindent
here $\beta=1/k_BT$, here $k_B$ Boltzmann constant which we consider to be 1, $V_{\rm acc}$ is the volume accessible to the pseudo neighbour, and $y(r)$ is a step function that takes into account that the potential is cut off at 2.5$\sigma_{\alpha \beta}$, i.e.~$y(r)=1$ if $L\leq r \leq L+2.5\sigma_{\alpha\beta}$ and $y(r)=0$ for all other values of $r$. The volume integrals in Eq.~\ref{prob_RP} can be decomposed into a spherical part that is contained inside the cubic box of length $L_{box}$ and the rest. The latter volume is given by, 
\begin{eqnarray}
\Delta V&=&L_{\rm box}^{3}-\frac{4}{3}\pi \Big(\frac{L_{\rm box}}{2}\Big)^{3}\\
&=&L_{\rm box}^{3}(1-\frac{\pi}{6}) \quad .
\label{del_V}
\end{eqnarray}
A spherical integration in Eq.~(\ref{prob_RP}) can be written as,
\begin{equation}
P_{RP}=\frac{\int_{L}^{L+r_c}dr\;4{\pi}r^{2}e^{-\beta u(r-L)}}{\int_{L}^{L_{\rm box}/2}dr\;4{\pi}r^{2}e^{-\beta u(r-L)}+\Delta V} \quad .
\label{prob_real_pseudo}
\end{equation}

Note that $P$ depends on the interaction potential via $u(r)$ and $r_c$. For a binary system, we can generalise this calculation to obtain the partial probabilities $P^{\alpha\beta}_{RP}$ and then the total probability is given by

\begin{equation}
P_{RP}=x_{A}^2P_{RP}^{AA}+2x_{A}x_{B}P_{RP}^{AB}+x_{B}^2P_{RP}^{BB}\quad ,
\label{total_prob_pseudo}
\end{equation}
\noindent
where $x_\alpha$ is the concentration of species $\alpha$.

\begin{figure}[h!]
 \centering
\vspace{0.5cm}
 \includegraphics[width=0.5\textwidth]{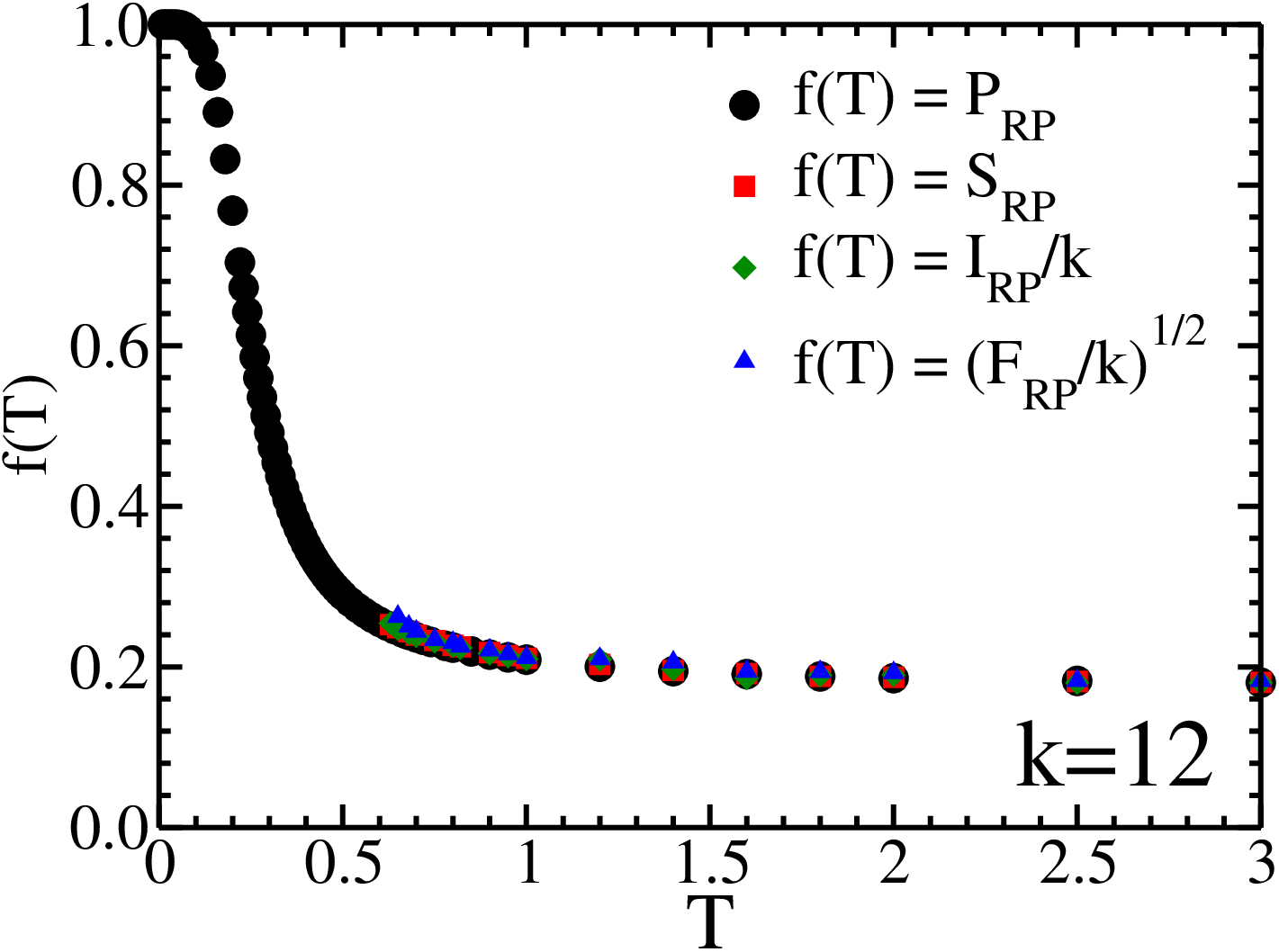}
 \caption{$P_{RP}$, the probability of finding a pseudo neighbour of a real particle, given by Eqs.\ref{prob_real_pseudo} and \ref{total_prob_pseudo}, is plotted against temperature for $k=12$ and  $L=2.5$ system. We also plot the saturation value of the real-pseudo bond breakage time correlation function,$S_{RP}$, and the initial and final number of real-pseudo bonds, $I_{RP}$ and $F_{RP}$, respectively for $k=12$ and $L=2.5$ system as obtained from simulation. Within the temperature range of the simulation we find that $P_{RP}=S_{RP}=I_{RP}/k= (F_{RP}/k)^{1/2}$.} 
 \label{P_k12}
 \end{figure}

Once we know the probability function $P_{RP}$, which describes the likelihood that a randomly assigned pseudo neighbour lies within the bonding cutoff distance, we can calculate the initial number of real–pseudo bonds per particle, given that there are $k$ assigned neighbours, as $I_{RP}=k*P_{RP}$. 
To estimate the number of bonds that persist over long times, the final bonds, we now need a conditional probability, the probability that a pseudo neighbour remains bonded at long times, given that it was bonded initially. This quantity can be expressed as $F_{RP}=I_{RP}*P_{RP}=k*P_{RP}^{2}$. Thus, the saturation value of the bond breakage time correlation function, $S_{RP}=\frac{F_{RP}}{I_{RP}}=P_{RP}$. In Fig.\ref{P_k12}, we plot $S_{RP}$, $I_{RP}/k$ and $(F_{RP}/k)^{1/2}$ as obtained from the simulation studies. We also plot $P_{RP}$ as obtained from  Eqs.\ref{prob_real_pseudo} and \ref{total_prob_pseudo}. We find that they overlap, which confirms our analysis. 

Next, we analyse the same for real-real bonds both in KA and the mean field system. Although the timescale of decay of the bonds in KA model and the real-real bonds in the mean field system are different (Fig. \ref{bond_break_KA} and Fig.\ref{real_real_bb}) since the local structures of the two systems given by the radial distribution function, $g(r)$ (defined in the Appendix\ref{overlap} and shown in Fig.\ref{rdff}) is same \cite{ujjwal_mean_field} this part of the analysis remains identical for both the systems. The probability, $P_{RR}$, that a real particle is near another real particle can be written when in Eq.\ref {prob_real_pseudo} we replace the interaction potential by the effective interaction between the particles,  $\it{i.e}$, the potential of mean force. Thus $\exp(- \beta u(r))$ in Eq.\ref{prob_real_pseudo} should be replaced by the radial distribution function $g(r)$. 

\begin{equation}
P_{RR}=\frac{\int_{0}^{r_{c}}dr\;4{\pi}r^{2}g(r)}{\int_{0}^{L_{\rm box}/2}dr\;4{\pi}r^{2}g(r)+\Delta V} \quad .
\label{prob_real_real}
\end{equation}

Similar to that for the real-pseudo bonds, for a binary system, we can generalise this calculation to obtain the partial probabilities $P^{\alpha\beta}_{RR}$ and then the total probability is given by

\begin{equation}
P_{RR}=x_{A}^2P_{RR}^{AA}+2x_{A}x_{B}P_{RR}^{AB}+x_{B}^2P_{RR}^{BB}\quad ,
\label{total_prob}
\end{equation}
\noindent
where $x_\alpha$ is the concentration of species $\alpha$.

\begin{figure}[h!]
 \centering
\vspace{0.5cm}
 \includegraphics[width=0.5\textwidth]{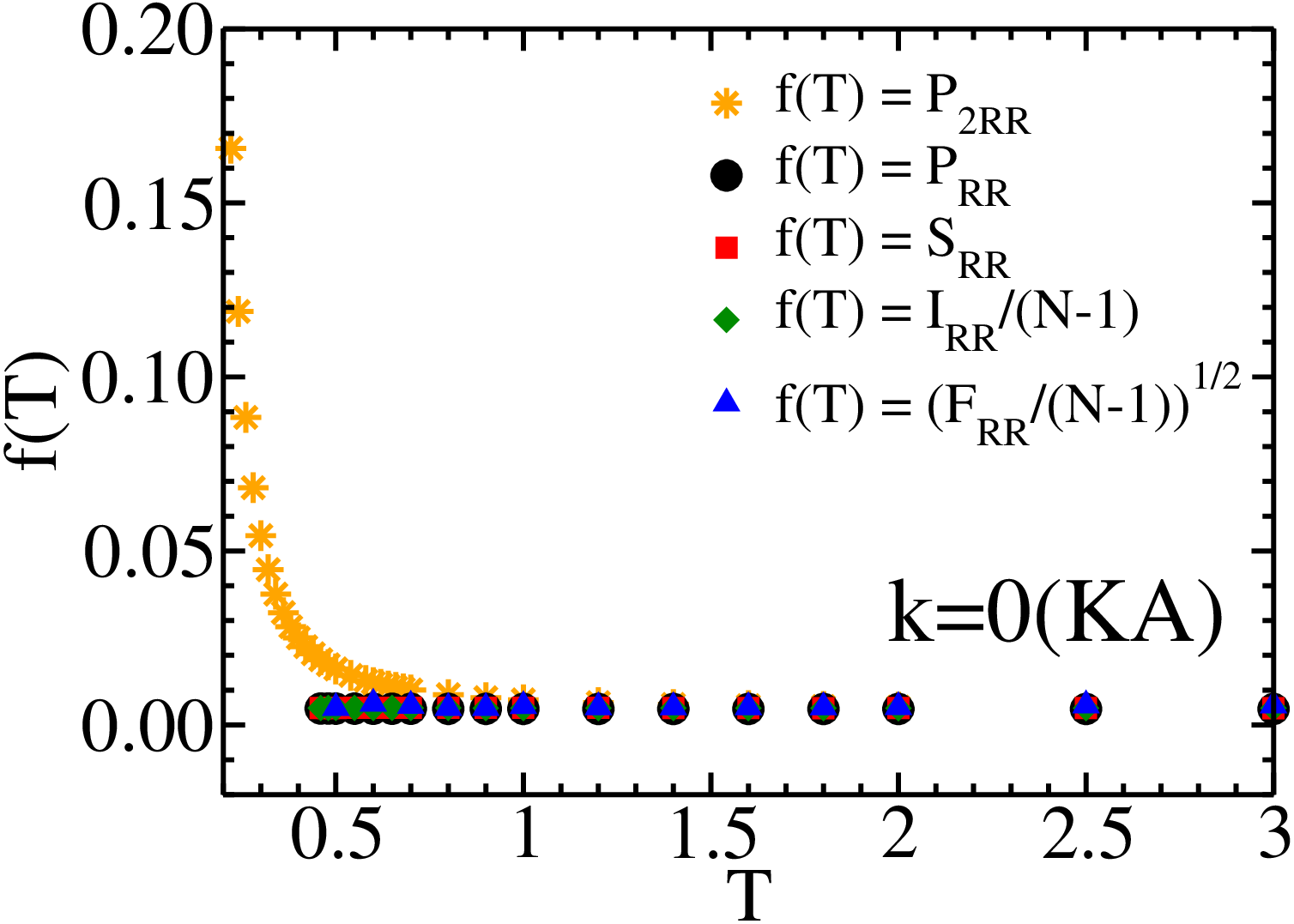}
 \caption{$P_{RR}$, the probability of finding a particle as a neighbour of another particle, given by Eqs.\ref{prob_real_real} and \ref{total_prob}, is plotted against temperature. We also plot the saturation value of the bond breakage time correlation function, $S_{RR}$, and the initial and final number of bonds, $I_{RR}$ and $F_{RR}$, respectively, as obtained from the simulation and we show that $P_{RR}=S_{RR}=I_{RR}/(N-1)=(F_{RR}/(N-1))^{1/2}$. We also plot $P_{2RR}$, which is the probability of finding a neighbour of a particle if the potential of mean force is equal to the interaction potential. This is obtained by replacing $g(r)$ in Eq.\ref{prob_real_real} by $\exp(- \beta u(r))$. We find that $P_{2RR}$ grows much more strongly with temperature compared to $P_{RR}$. This implies that higher-order correlations present in $g(r)$ facilitate bond breaking.  Note that this analysis is common for the real-real bonds in both the mean field system and that in the parent KA (k=0) system.}
 \label{P_for_KA}
 \end{figure}
 \noindent

Once we know the probability $P_{RR}$, which describes the likelihood that a particle will lie within the bonding cutoff distance of another particle, we 
can calculate the initial number of bonds per particle within the distance $r_{c}=1.4 \sigma_{11}$, given that any of the remaining (N-1) particles can be the neighbour of the central particle. This initial number of bonds is given by $I_{RR}=(N-1)*P_{RR}$. The number of initial bonds that survive after a long time is again a conditional probability, and the final number of bonds can be written as $F_{RR}=I_{RR}*P_{RR}=(N-1)*P^{2}_{RR}$. Thus the saturation value is given by $S_{RR}=\frac{F_{RR}}{I_{RR}}=P_{RR}$. In Fig.\ref{P_for_KA} we plot $S_{RR}$,$I_{RR}/(N-1)$ and $(F_{RR}/(N-1))^{1/2}$ as obtained from simulations for the real-real bonds in the mean field system. We then plot $P_{RR}$ as obtained from Eqs.\ref{prob_real_real} and \ref{total_prob}. We find that the functions overlap, confirming our analysis. We also plot the probability of finding particles $P_{2RR}$ where the probability is calculated by replacing $g(r)$ in Eq.\ref{prob_real_real} by $\exp(- \beta u(r))$. Thus, $P_{2RR}$ provides the probability of finding particles if all the particles are interacting only via two body interactions. As observed from the plot, at high temperatures $P_{2RR} \simeq P_{RR}$. However at low temperatures $P_{2RR}$ grows faster than $P_{RR}$. This implies that at low temperatures, many-body interactions lower the probability of an initial bond to survive. This is because once a bond is broken and a particle moves a little away, the position is taken up by another neighbouring particle. This lowers the probability of the particle bonding back to the central particle. Note that due to the low density of the pseudo neighbours, the pseudo interaction, as shown earlier, is primarily two body in nature \cite{ujjwal_mean_field}. This is precisely the reason why we find a strong temperature dependence in the $S_{RP}$, and a strong probability of rebonding, which leads to residual pseudo bonds which grow strongly in number at low temperatures.

\subsection{System size effect of the bond breakage dynamics}
Note that in Eq.\ref {prob_real_pseudo} and Eq.\ref {prob_real_real} the denominator is essentially the total volume of the system. This predicts that the saturation values, $S_{KA}$, $S_{RR}$ and $S_{RP}$ will reduce with the increase in the system size. Since the number of surviving bonds is related to the saturation value, there is a chance that in the limit of infinite system size, these surviving bonds will disappear, and the effect of them that we find here is a system size effect. Thus, we study the system size effect on the bond breakage time correlation functions.

\label{}
\begin{figure}[h!]
 \centering
\vspace{0.5cm}
 \includegraphics[width=0.5\textwidth]{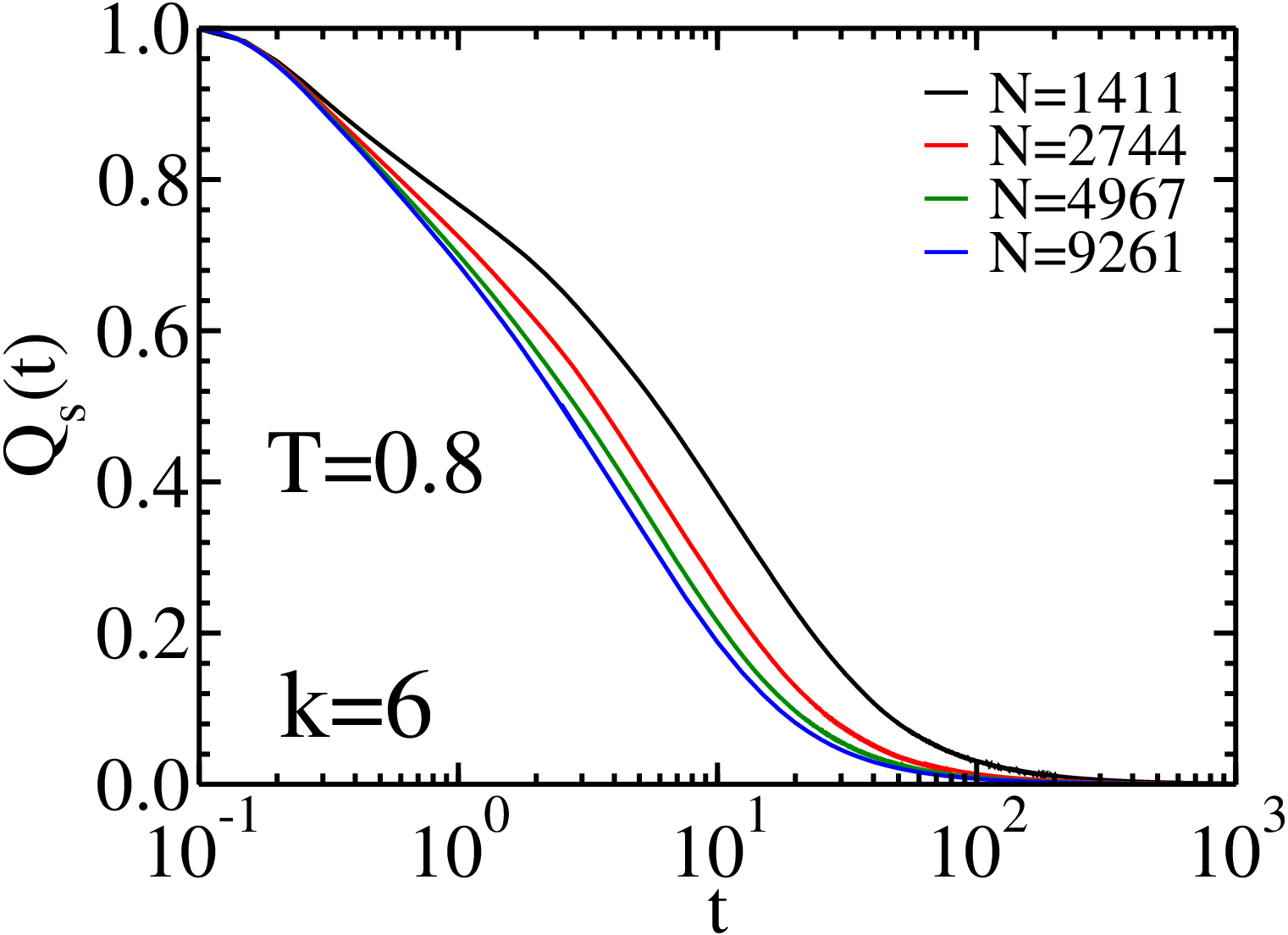}
 \caption {System size effect of the self part of the overlap function at $T=0.8$ and $k=6$ with system sizes $N=1411$ ($\rho_k=0.0054$), $N=2744$ ($\rho_k=0.0026$), $N=4967$ ($\rho_k=0.0015$) and $N=9261$ ($\rho_k=0.0008$). $\rho_k$ is the pseudo neighbour density of the mean field system. We find that with an increase in system size, while keeping $k$ constant, the $\rho_{k}$ decreases and the dynamics becomes faster.} 
 \label{diff_density_overlap}
 \end{figure}

 \begin{figure}[h!]
 \centering
 \includegraphics[width=0.5\textwidth]{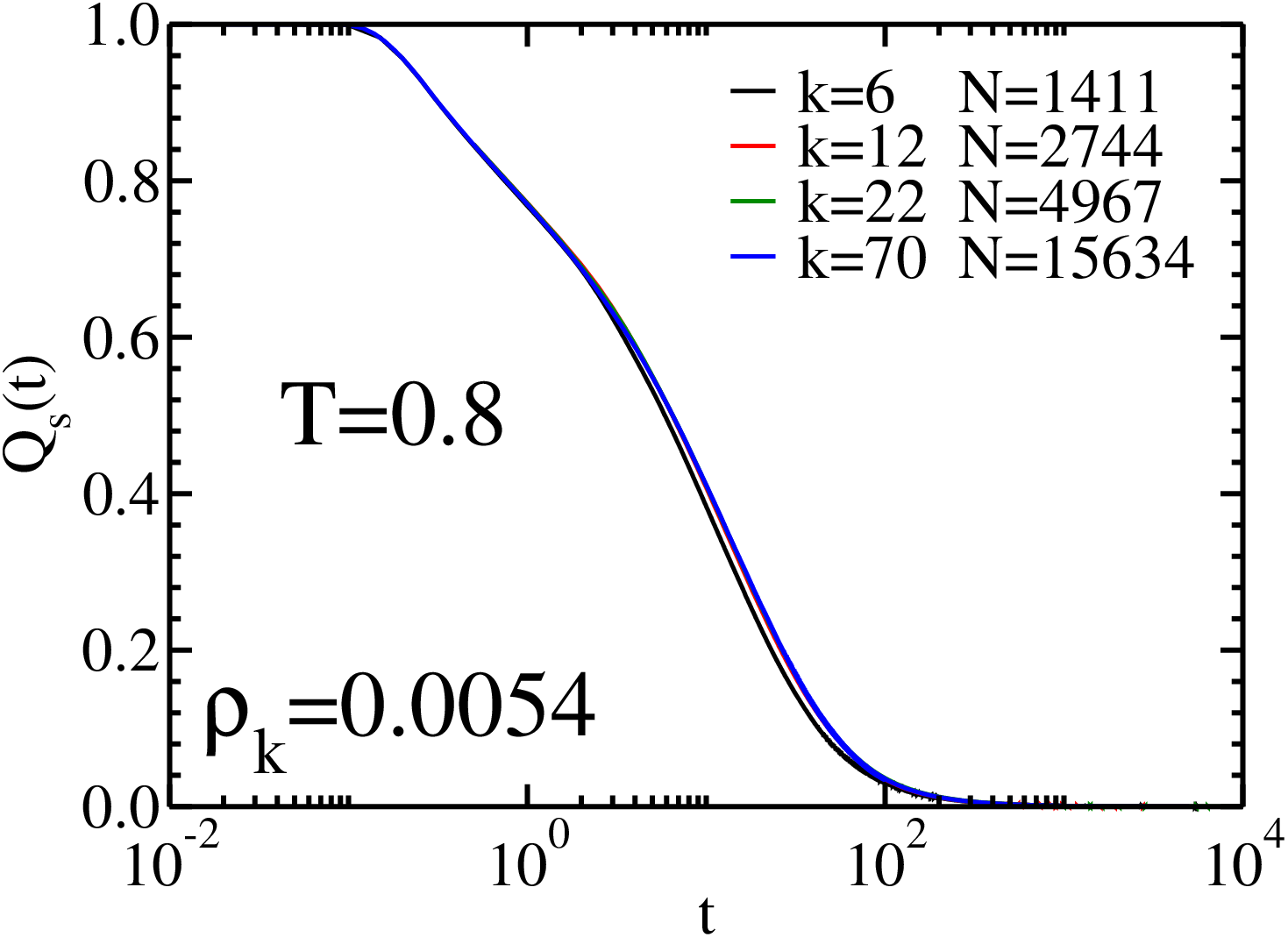}
 \caption{System size effect on the self part of the overlap function at $T=0.8$ and $\rho_k=0.0054$. Here, keeping $\rho_{k}$ constant, the number of pseudo neighbours is also increased with the system size. We find that at constant $\rho_{k}$, the dynamics is similar.}
 \label{same_density_overlap}
 \end{figure}

While studying systems of different sizes, we find that if we keep the number of pseudo neighbours the same and increase the system size, then the dynamics of the system becomes faster (Fig.\ref{diff_density_overlap}). This is because with an increase in the system size, the density of the pseudo neighbours, $\rho_{k}=\frac{k}{V-\frac{4}{3}\pi L^{3}}$ reduces and since the dynamics is dependent on $\rho_{k}$ it becomes faster. Thus, in order to study equivalent systems with similar dynamics, we need to keep $\rho_{k}$ fixed (Fig.\ref{same_density_overlap}). Hence, along with the system size, we also increase the number of pseudo neighbours.
Thus, we create a few systems with N=1411 and k=6, N=2744 and k=12, N=4967 and k=22 and N=15634 and k=70. Note that for all these systems the $\rho_{k}=0.0054$ and L=2.5. For the same system sizes, we also study the bond breakage dynamics in a KA model.  

\begin{figure}[h!]
 \centering
 \includegraphics[width=0.5\textwidth]{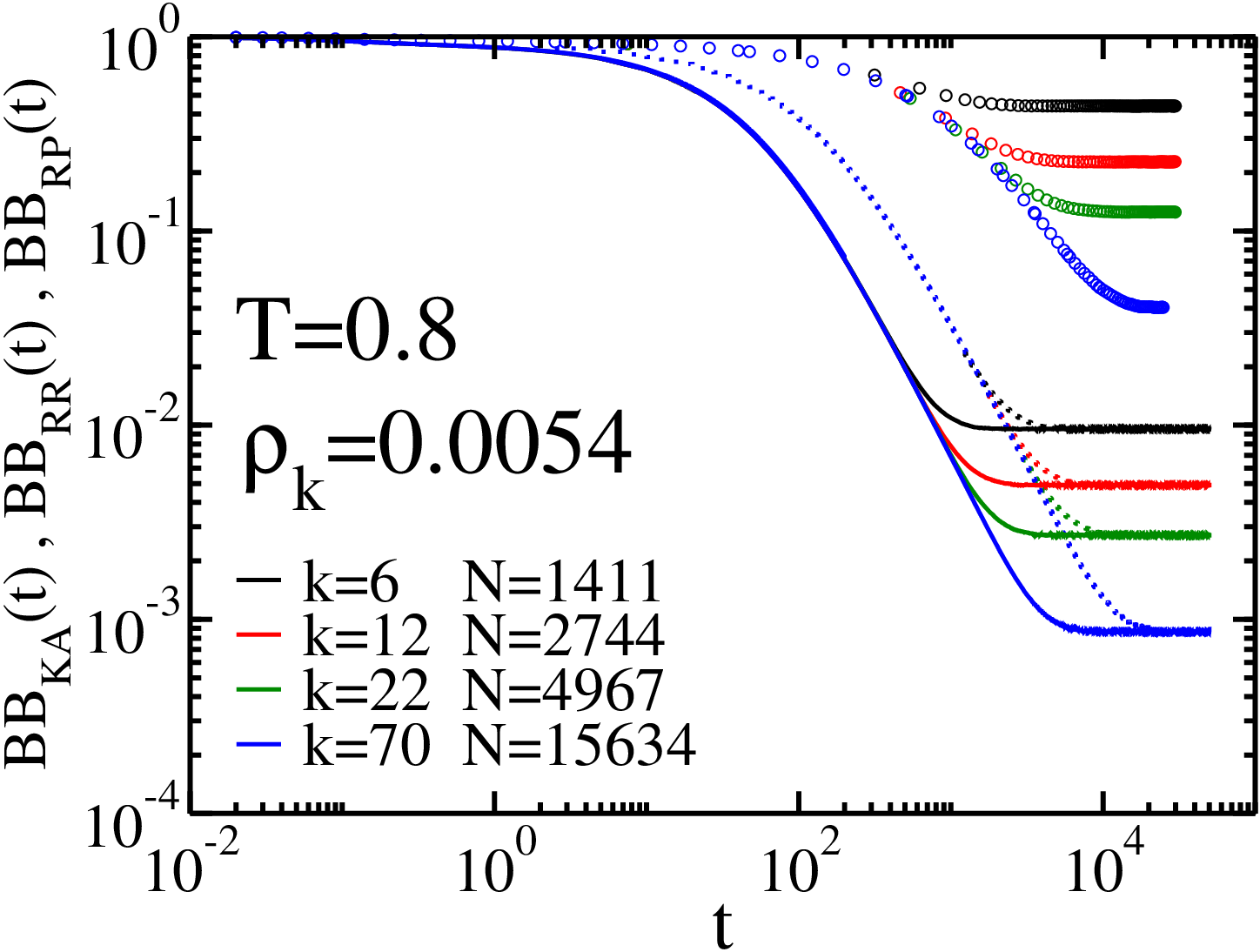}
 \caption{System size effect on the time evolution of the bond breakage time correlation functions. $BB_{KA}(t)$, bond breakage time correlation function, for $KA$ (k=0) system (solid lines), $BB_{RR}(t)$, real-real bond breakage time correlation function for mean field system  (dotted lines) and $BB_{RP}(t)$, real-pseudo bond breakage time correlation function (symbols) for mean field system at $T=0.8$ and pseudo neighbour density $\rho_k=0.0054$ but for different system sizes. For all three types of correlation functions, the saturation value decreases with system size; however, the curves of the same function overlap at shorter times. Thus, in this temperature regime, apart from the saturation value, the dynamics is independent of system size and dependent on $\rho_{k}$.} 
 \label{comp_BB_system_size}
 \end{figure}

\begin{figure}[h!]
 \centering
\vspace{0.5cm}
 \includegraphics[width=0.5\textwidth]{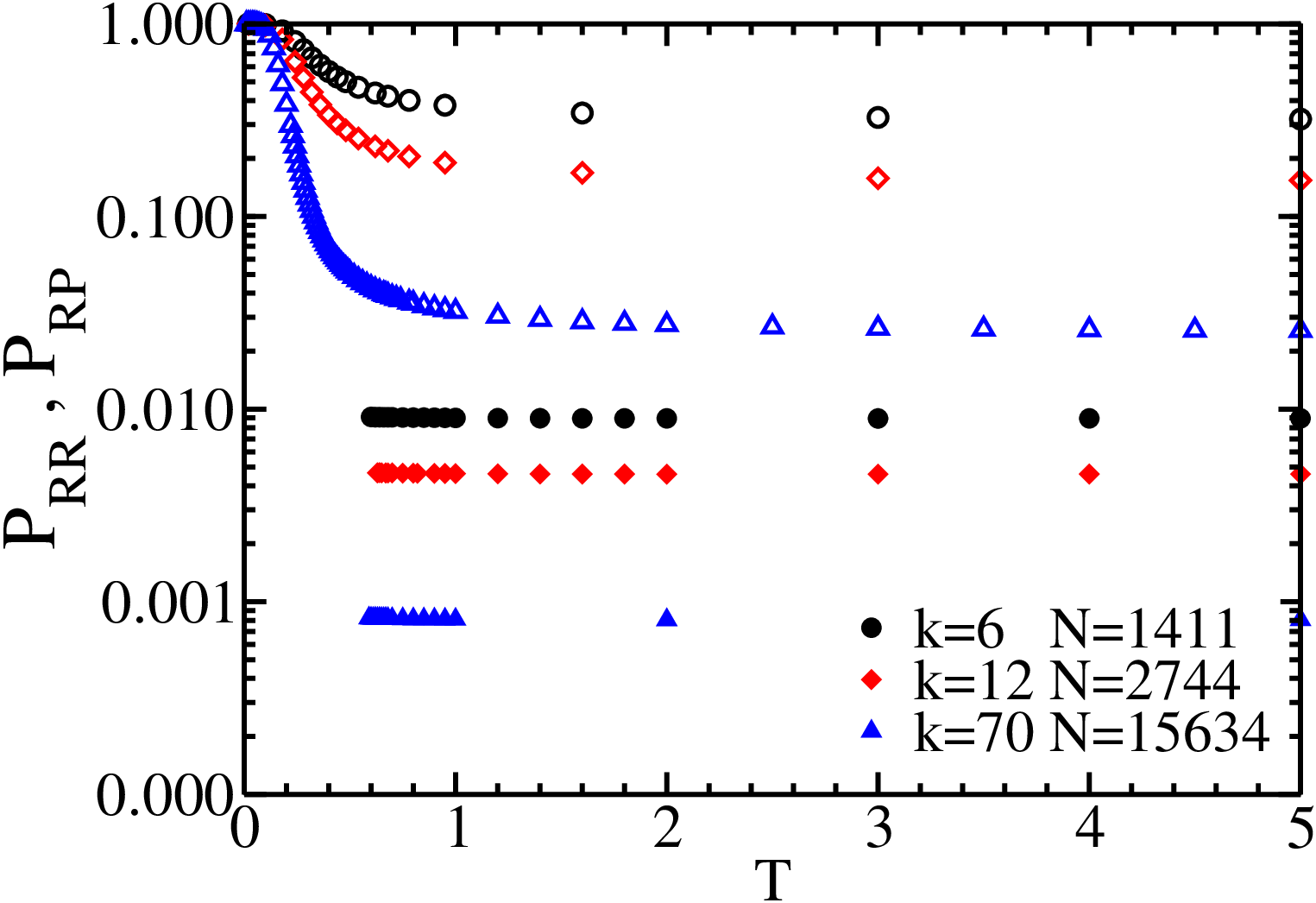}
 \caption{System size effect of the probability:. $P_{RR}$,(filled symbols)  probability of finding a real neighbour of a real particle both in the mean field and KA (k=0) systems and $P_{RP}$, (open symbols) the probability of finding a pseudo neighbour of a real particle in the mean field system plotted as a function of temperature. The systems are chosen such that $\rho_{k}=0.0054$ is constant. Similar to that we find in the saturation value of the bond breakage correlation, the probability of finding a neighbour decreases with system size.}
 \label{prob_system_size}
 \end{figure}

\begin{figure}[h!]
 \centering
 \includegraphics[width=0.5\textwidth]{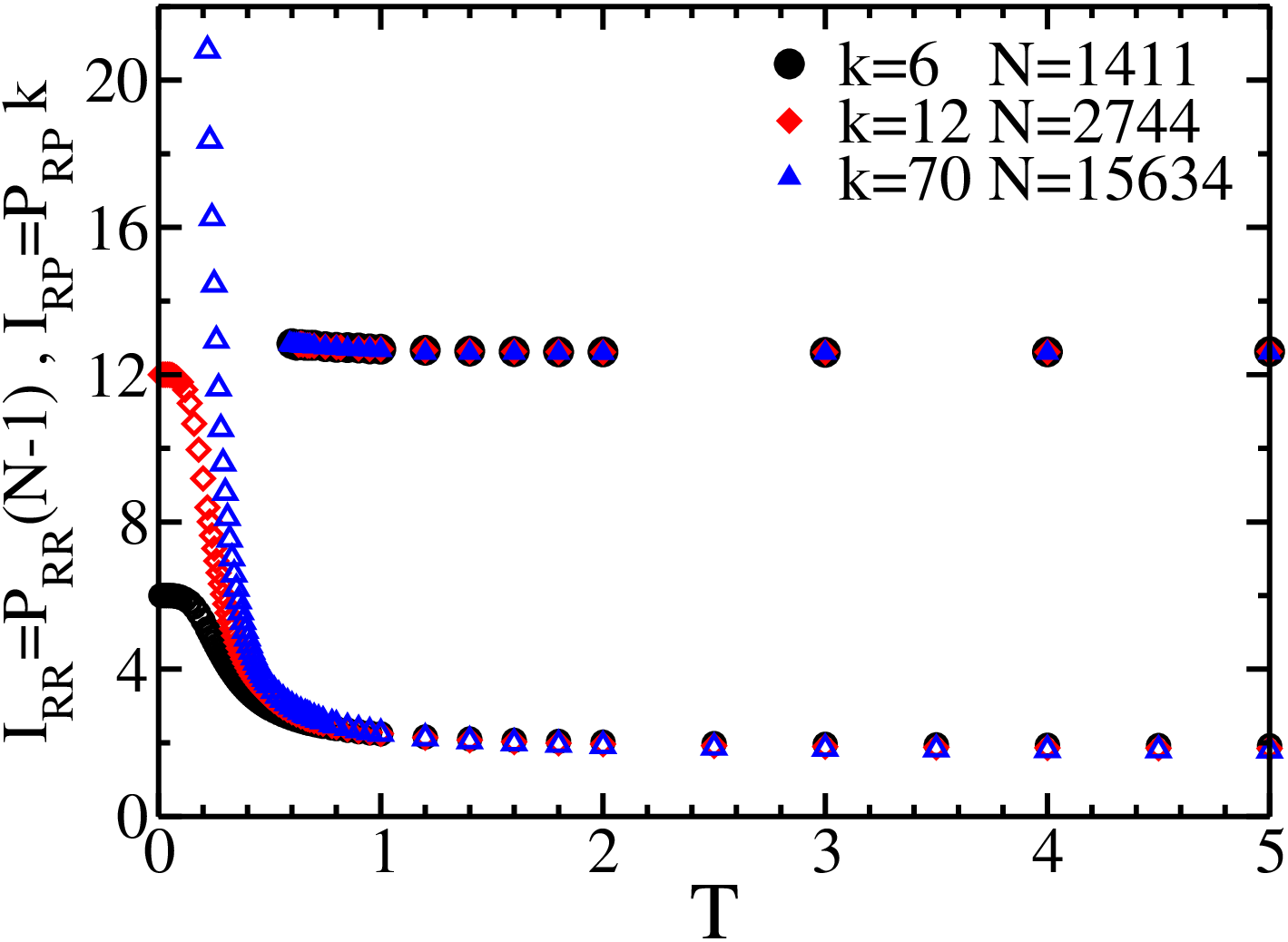}
 \caption{System size effect of initial bonds: $I_{RR}$, (filled symbols), the initial number of real-real bonds both in the mean field system and KA (k=0) systems and $I_{RP}$, (open symbols) the initial number of real-pseudo bonds in the mean field system plotted as a function of temperature. The systems are chosen such that $\rho_{k}=0.0054$ is constant. The number of real-real initial bonds is independent of the system size. The number of real-pseudo initial bonds at high temperatures is independent of the system size, but at low temperatures, it grows with system size.}
 \label{initial_system_size}
 \end{figure}

\begin{figure}[h!]
 \centering
 \includegraphics[width=0.5\textwidth]{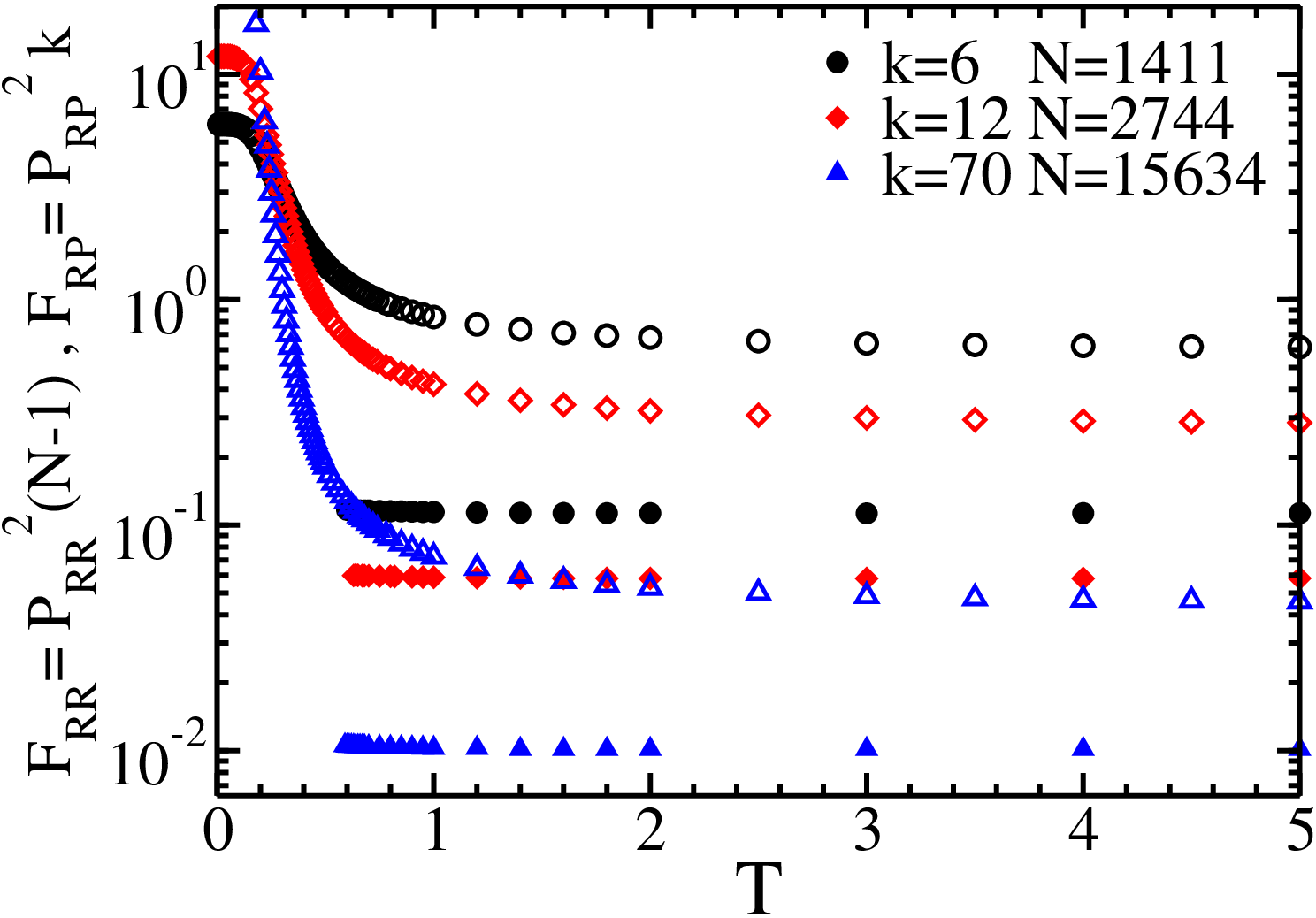}
 \caption{System size effect of surviving final bonds: $F_{RR}$, (filled symbols) the final number of real-real bonds both in the mean field system and KA (k=0) systems and $F_{RP}$ (open symbols) the final number of real-pseudo bonds in the mean field system plotted as a function of temperature. The systems are chosen such that $\rho_{k}=0.0054$ is constant. The real-real final number of bonds decreases with increasing system size. The real-pseudo final number of bonds at high temperatures decreases with the increasing system size, but at low temperatures, it grows with system size. } 
 \label{final_system_size}
 \end{figure}

In Fig.\ref{comp_BB_system_size} we plot the bond breakage dynamics at a particular temperature for the KA system, real-real neighbours and real-pseudo bonds in the mean field system for different system sizes. We find that for all three, the timescale of relaxation is independent of system size, but as expected, the saturation value is system size dependent. It decreases with increasing system size. The rest of the features are the same as discussed for Fig.\ref{comp_BB}. 
In the previous section, we showed that the initial bonds, the final bonds and also the saturation values for both KA and mean field systems can all be calculated from the probability functions (Eq.\ref{total_prob_pseudo} and Eq.\ref{total_prob}). This allows us to access temperature regimes which are not accessible in simulation studies. Thus, to understand the system size effect over a wide temperature range, we use the analytical formulas. The saturation values, $P_{RR}$ and $P_{RP}$ at all temperatures are plotted in Fig.\ref{prob_system_size}. Note that $P_{RR}$ describes both the KA system and the real neighbours in the mean field system. We find that the $P_{RR}$ value decreases with system size, and the system size effect is the same at all temperatures. For $P_{RP}$, the system size effect at high temperatures shows a similar trend to that for $P_{RR}$. However, at low temperatures, the difference between the curves starts to decrease. In Fig.\ref{initial_system_size} we plot $I_{RR}=(N-1)*P_{RR}$, (valid for both the KA systems and real bonds in the mean field systems). We also plot $I_{RP}=k*P_{RP}$ for the mean field systems. We find that $I_{RR}$ is constant and independent of the system size. This is expected as the number of real neighbours should not depend on the system size. However, for the mean field system, although at high temperatures the number of pseudo neighbours (initial pseudo bonds) is independent of the system size, at low temperatures the number of pseudo neighbours (bonds) grows strongly with a decrease in temperature. The growth rate is higher for larger system sizes, as eventually at $T=0$, $P_{RP}=1$ and $I_{RP}=k$ and for a constant $\rho_{k}$, $k$ increases with system size.  Next, we study the number of final surviving bonds (Fig.\ref{final_system_size}). We find that $F_{RR}$ decreases with increasing system size. However, $F_{RP}$, although at high temperature shows a similar decrease with increasing system size, at low temperatures the scenario reverses. We can make a few important observations from the plots in Fig.\ref{initial_system_size} and Fig.\ref{final_system_size}.
i) This analysis is presented for a system with a moderately low pseudo neighbour density, $\rho_{k}=0.0054$. In this regime, we observe that, independent of system size, the number of initial real bonds is more than twice the number of initial pseudo bonds. However, across all temperature ranges studied, the number of surviving pseudo bonds is at least an order of magnitude higher than that of the surviving real bonds. ii) Low numbers of surviving real bonds have negligible influence on the dynamics, and this influence further diminishes with system size. In contrast, surviving pseudo bonds exert a significant and growing influence on the dynamics, particularly at low temperatures and larger system sizes. This highlights a key difference: while the bond breakage dynamics of real bonds follow the overall relaxation dynamics of the system, the pseudo bond dynamics decouple from them. Since the bond breakage correlation function for pseudo bonds does not decay to zero or to a temperature independent plateau, it is not possible to define a characteristic timescale from this function.  However, in the following section, we investigate the correlation between the number of surviving pseudo bonds, which are known to affect bond breakage dynamics and the thermodynamic transition temperature $T_{K}$.

\begin{figure}[h!]
 \centering
 \includegraphics[width=0.5\textwidth]{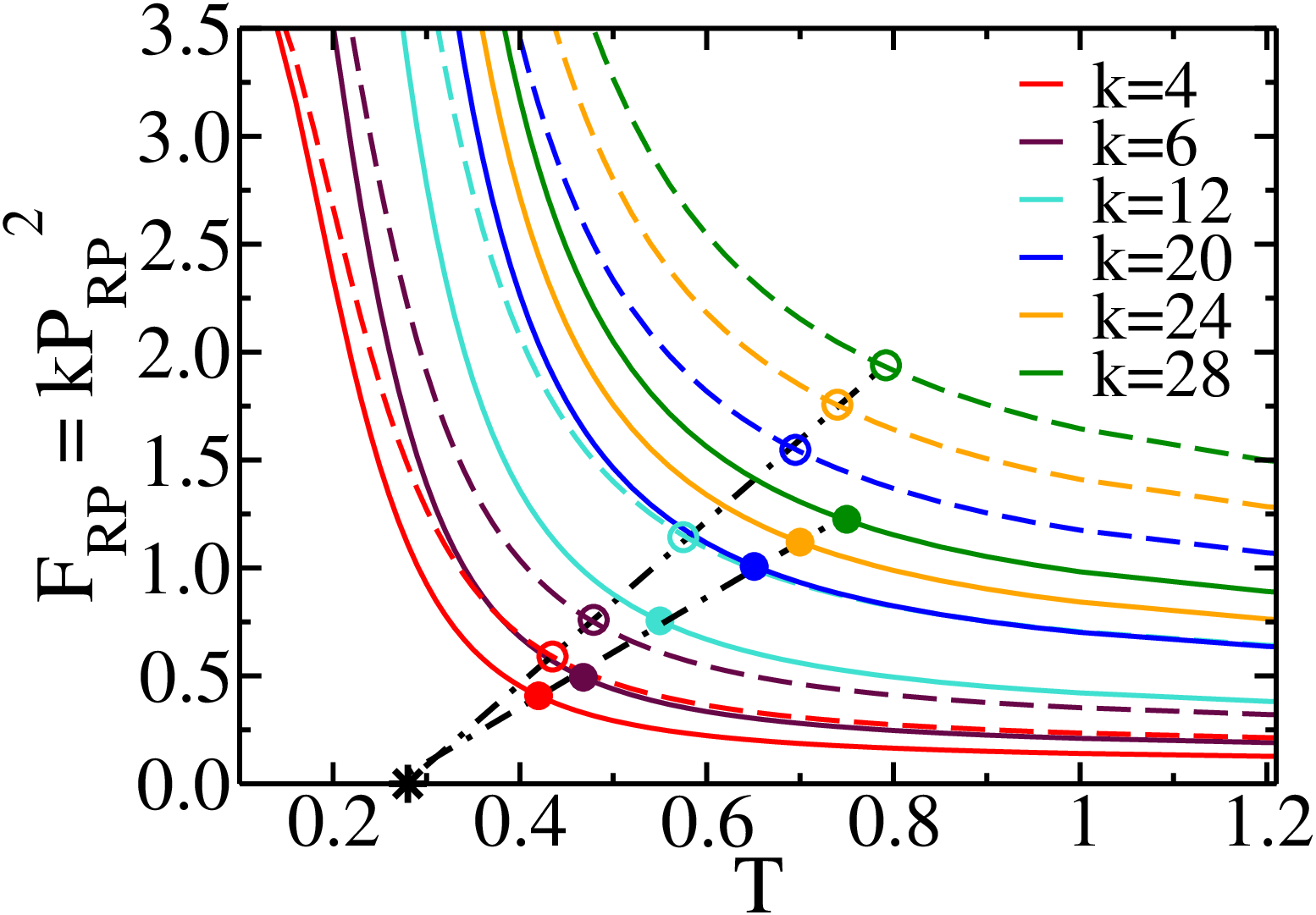}
 \caption{Final number of surviving pseudo bonds, $F_{RP}$ as a function of temperature for systems with pseudo neighbours, $k=12,20,24$ and $28$. Dashed lines are for $L=3.0$, and solid lines are for $L=2.5$. Open symbols are $F_{RP}(T_K)$ values for $L=3.0$, and filled symbols are $F_{RP}({T_K})$ values for $L=2.5$. In the same plot, we also mark the $T_{K}=0.28$ for the KA  ($k=0$) system on the x-axis (black star), as for this system $F_{RP}=0$. Each of the dashed dotted lines connect the $F_{RP}(T_{k})$ values of the different systems with constant $L$.} 
 \label{final_bond_tk}
 \end{figure}
 
\begin{figure}[h!]
 \centering
 \includegraphics[width=0.5\textwidth]{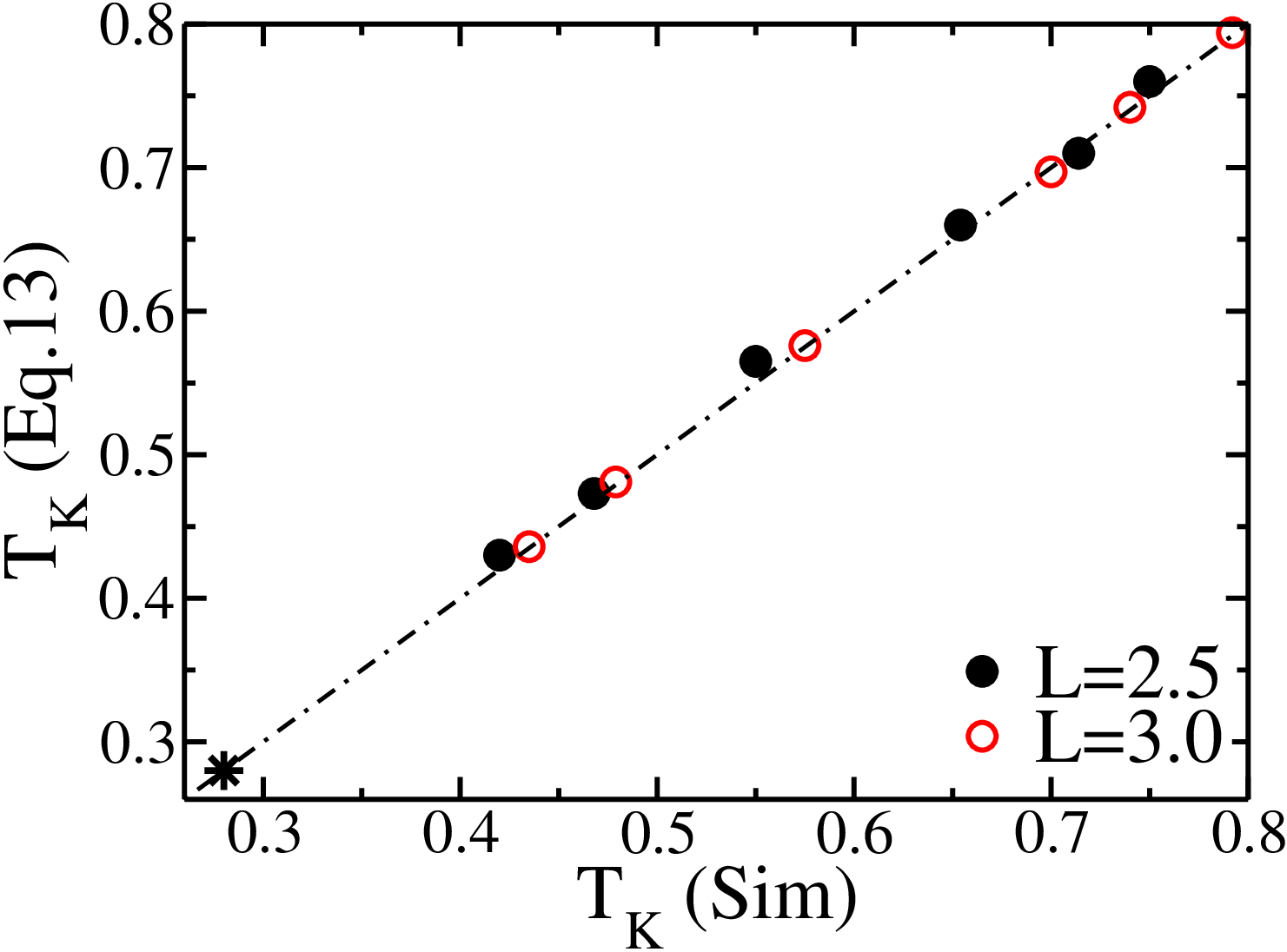}
 \caption{The Kauzmann temperature, $T_K$ obtained from simulation versus the Kauzmann temperature, $T_K$ calculated from the analytical expression Eq. \ref{k_dependence_of_TK}. The points fall very close to the y=x line shown by the dashed dotted line. This shows the validity of  Eq.\ref{k_dependence_of_TK}. The star symbol represents $T_K$=0.28 for $KA$ (k=0) system.}
 \label{tk_vs_tk}
 \end{figure} 
 
 \begin{figure}[h!]
 \centering
 \includegraphics[width=0.5\textwidth]{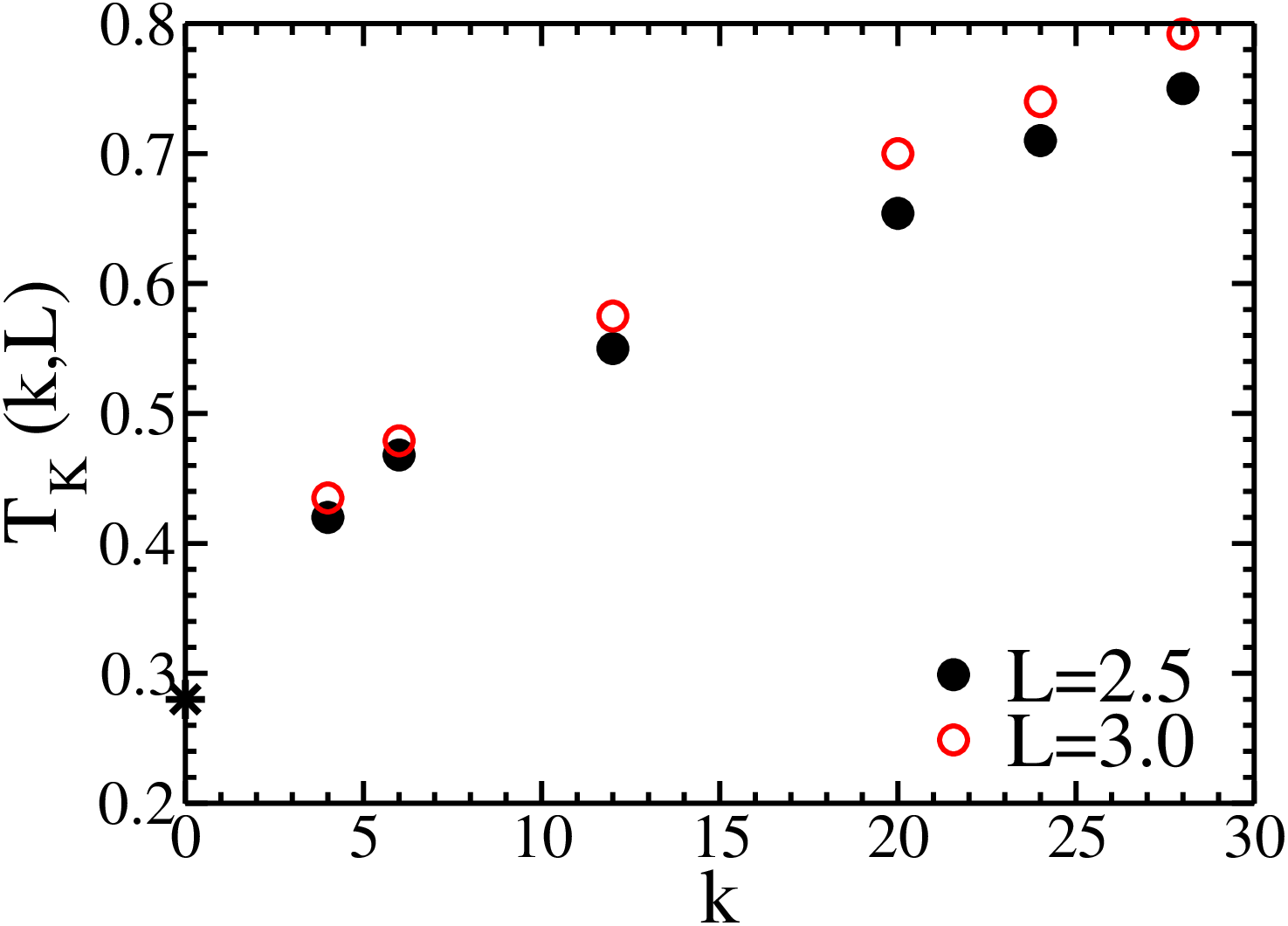}
 \caption{Kauzmann temperature, $T_K$, versus $k$ for $L=2.5$ and $L=3.0$. The star symbol represents $T_K$=0.28 for $KA$ (k=0) system. It shows that although $T_{K}$ depends on $k$ (Eq.
\ref{k_dependence_of_TK}), due to the presence of $P^2(L,V,T_K)$ the dependence is slightly weaker than linearity.} 
 \label{k_vs_tk}
 \end{figure}

 \subsection{Thermodynamics and its relation with the number of surviving bonds}

Here we explore the relationship between the final surviving pseudo bonds, $F_{RP}$ and the thermodynamic transition at $T_{K}$.
The analysis is done exploiting the fact that we can calculate the final number of bonds from the analytical expressions of $P_{RP}$. This allows us to explore temperature ranges near and below $T_{K}$. 
In Fig.\ref{final_bond_tk}, we plot the $F_{RP}=k*P^{2}_{RP}$ as obtained from Eq.\ref{prob_real_pseudo} and Eq.\ref{total_prob_pseudo} against temperature for different $k$ values for two different sets of systems ($L=2.5$ and $L=3.0$). We next show the value of $F_{RP}$ at $T_{K}$ by closed (for $L=2.5$) and open (for $L=3.0$) circles. In the same figure, on the x-axis, we also mark the $T_{K}$ value for the KA (k=0) system. Note that for the KA system $F_{RP}=0$. We find that the final number of bonds at $T_{K}$ is system dependent. However, there appears to be a pattern. For systems with more pseudo neighbours the $T_{K}$ appears at a higher $F_{RP}$ value. From  Fig.\ref{final_bond_tk}, we can write the following phenomenological expression relating the $T_{K}$ to the final number of bonds,

\begin{eqnarray}
{T_K}(k,L)&=& \alpha_{L}F_{RP}(k,L,T_{K}) + T_{K}(k=0)\nonumber\\
&=&\alpha_{L} kP_{RP}^2(L,T_K) + T_K(k=0)
\label{k_dependence_of_TK}
\end{eqnarray}
\noindent
here $T_{K}(k=0)=0.28$ is the Kauzmann temperature for the pure KA system. $\alpha_{L}$ is the inverse slope of the dashed-dotted line, which connects the $F_{RP}$ values at $T_{K}$ for the systems with different $k$ but same $L$ values (Fig.\ref{final_bond_tk}). From Eq.\ref{k_dependence_of_TK} we find that $T_{K}$ is proportional to $k$ and $P_{RP}^2(L,T_K)$ where $\alpha_{L}$ is the proportionality constant which has a weak $L$ dependence (Fig.\ref{final_bond_tk}).  Note that $F_{RP}(k,L)$ increases and $\alpha_{L}$ decreases with $L$. The increase in the former is greater than the decrease in the latter, and this leads to the increase of $T_{K}$ with $L$. Thus, the effect of the pseudo neighbours on the thermodynamics of the system increases with $L$, which is expected as with an increase in $L$, the pseudo neighbour density $\rho_{k}$ increases. To test the accuracy of the expression, in Fig.\ref{tk_vs_tk}, we plot $T_{K}$ calculated using Eq.\ref{k_dependence_of_TK} against $T_{K}$ obtained from simulation for all the systems. We find that all the data almost fall on the straight line with no offset and slope 1 (y=x).

From Eq.\ref{k_dependence_of_TK} we find that as $k$ increases, so does $T_{K}$, but the dependence is weaker than linearity because $P_{RP}^2(L,T)$ decreases with increasing temperature. This can be seen in the $T_{K}$ vs $k$ plot in Fig\ref{k_vs_tk}. We also find that as $k$ increases, the effect of $L$ becomes stronger and the difference between the $T_{K}$ values of two systems with the same $k$ but different $L$ increases.

\begin{table}[h!]
 \centering
 \caption{The values of characteristic temperatures for mean field systems with different $k$ values at two different $L=2.5$ and $L=3.0$. $T_K(Sim)$ is the Kauzmann temperature estimated from simulation and $T_K(Eq.\ref{k_dependence_of_TK})$ is the Kauzmann temperature estimated from Eq.\ref{k_dependence_of_TK}}
\begin{tabular}{|c|c|c|c|c|}
%\hline
%System& $L=2.5$&   & $L=3.0$ & &  \\
%\hline 
\hline
System &  \multicolumn{2}{|c|}{L=2.5} & \multicolumn{2}{|c|}{L=3}\\
\cline{2-5}
 & $T_K(Sim)$ & $T_K(Eq.\ref{k_dependence_of_TK})$ & $T_K(Sim)$ & $T_K(Eq.\ref{k_dependence_of_TK})$\\\hline
%\cline{1-5}
$k=4$ & 0.42 & 0.43 &  0.435 & 0.436 \\\hline
%\cline{1-5}
$k=6$ & 0.468 & 0.473 &  0.479 & 0.481 \\\hline
%\cline{1-5}
$k=12$ & 0.55 & 0.565 &  0.575 & 0.576 \\\hline
%\cline{1-5}
$k=20$ & 0.654 & 0.66 &  0.70 & 0.697 \\\hline
%\cline{1-5}
$k=24$ & 0.714 & 0.71 &  0.74 & 0.742 \\\hline
%\cline{1-5}
$k=28$ & 0.75 & 0.76 &  0.792 & 0.794 \\\hline
%\cline{1-5}
 \end{tabular}
  \label{Tk_table}
\end{table}

We make an interesting observation from the phenomenological expression written down by us, Eq.\ref{k_dependence_of_TK}. We find that in the mean field system, the contributions of real neighbours and pseudo neighbours in the determination of $T_{K}$ appear in an additive manner. The real neighbours act in a similar way as the pure KA system, and thus the contribution of real neighbours in $T_{K}$ is given by $T_{K}(k=0)$. We can then quantify the increase of $T_{K}$, $\Delta T_{K}$ due to the presence of the pseudo neighbours as,
\begin{eqnarray}
\Delta T_{K}={T_K}(k,L)-T_K(k=0)&=&\alpha_{L} F_{RP}(L,T_K) \nonumber\\
&=&\alpha_{L} kP_{RP}^2(L,T_K) \label{delta_Tk}
\end{eqnarray}

Thus, the increase in $T_{K}$ of the mean field system from that of the pure KA system can be related directly to the number of surviving pseudo bonds. This suggests that these surviving pseudo bonds are well correlated with the change in the thermodynamics of the system. However, as seen from the earlier analysis, these surviving pseudo bonds have a weak effect on the other dynamics, like the self and collective overlap function of the system. Because for these functions to decay, the breaking of the real-real bonds is good enough. This leads to the breakdown of the Adam Gibbs relationship. 

 \begin{figure}[h!]
 \centering
 \includegraphics[width=0.5\textwidth]{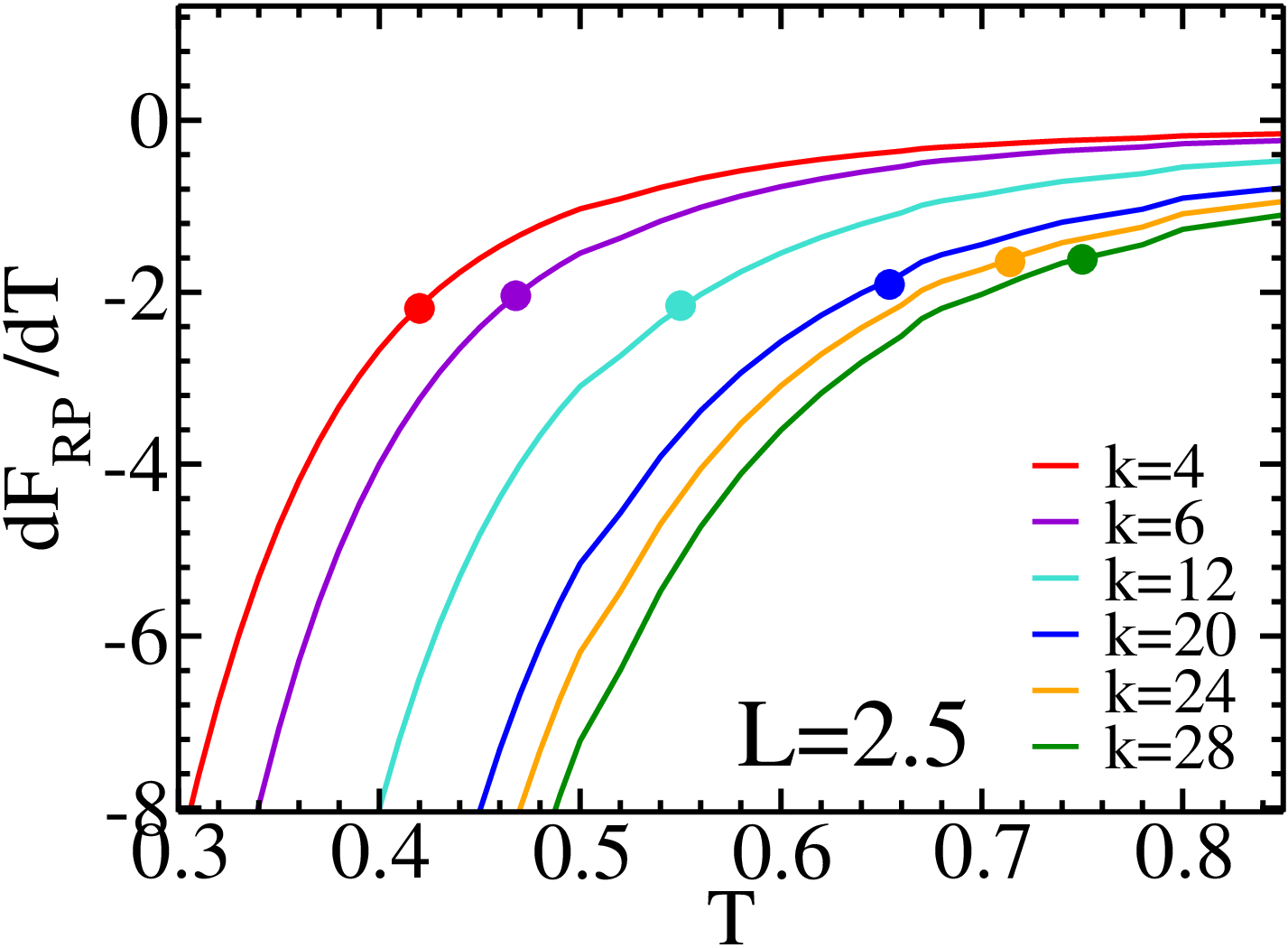}
 \caption{The temperature derivative of the final number of bonds, $\frac{dF_{RP}}{dT}$ as a function of temperature. The filled circles are $\frac{dF_{RP}}{dT}(T=T_K)$ values at the corresponding $T_{K}$ values for $k=12,20,24$, and $28$ systems at $L=2.5$} 
 \label{derv_final_bond_2.5}
 \end{figure}

\begin{figure}[h!]
 \centering
\vspace{0.5cm}
 \includegraphics[width=0.5\textwidth]{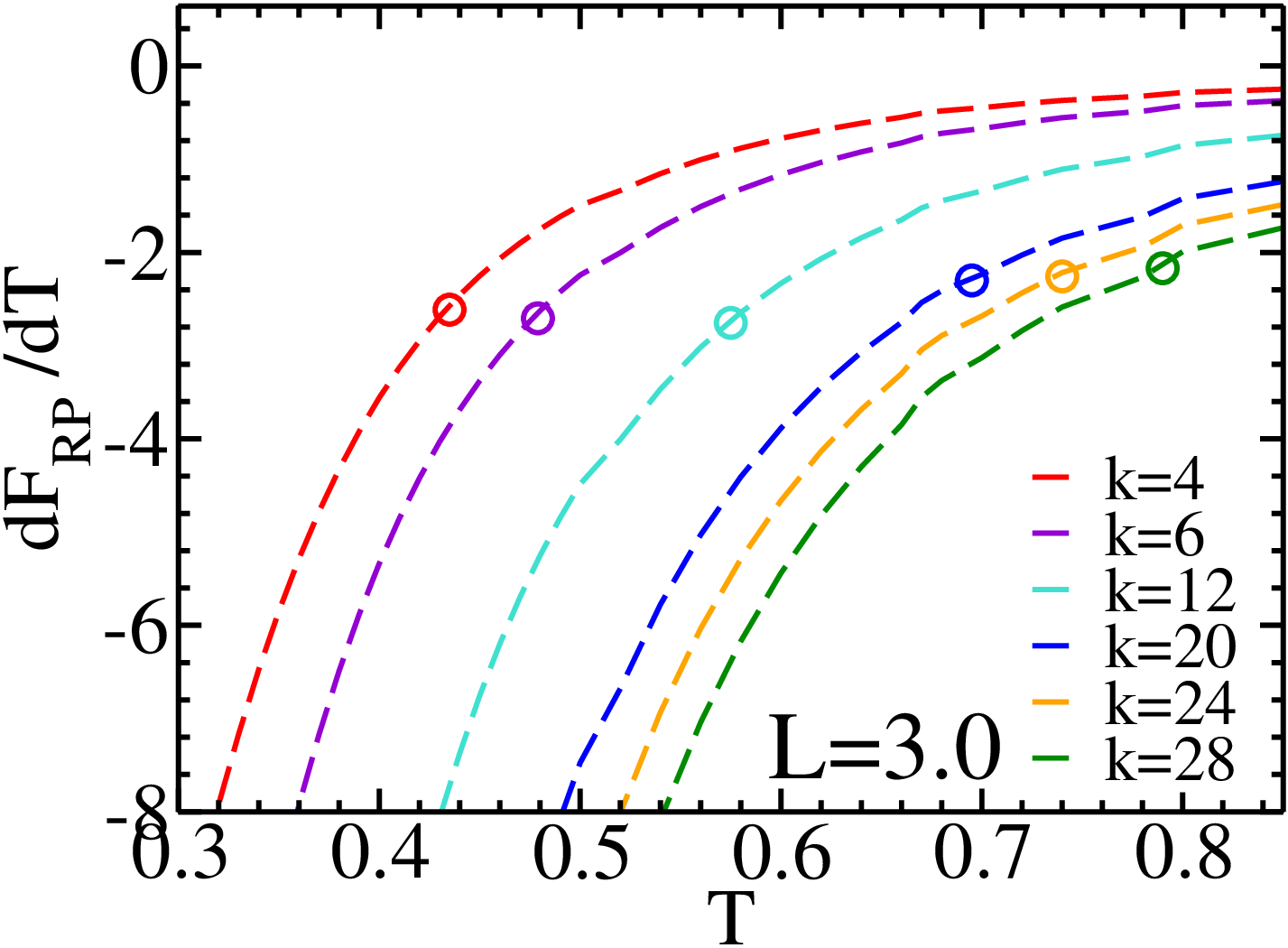}
 \caption{The temperature derivative of the final number of bonds, $\frac{dF_{RP}}{dT}$ as a function of temperature. The open circles are $\frac{dF_{RP}}{dT}(T=T_K)$ values at the corresponding $T_{K}$ values for $k=12,20,24$, and $28$ at $L=3.0$}
 \label{derv_final_bond_3.0}
 \end{figure} 
As discussed earlier, the thermodynamic transition at $T_{K}$ does not appear to occur at a constant value of $F_{RP}$. If we take a derivative of the $F_{RP}$ in terms of T, we find that all the $T_{K}$ values fall in a near horizontal line (Figs \ref{derv_final_bond_2.5} and \ref{derv_final_bond_3.0}).
Note that the number of final bonds at high temperatures is almost constant, and thus, the growth rate of final bonds is close to zero. However, with a decrease in temperature, they suddenly grow. It appears that when the growth rate of $F_{RP}$ reaches some finite value, the system undergoes a thermodynamic transition. At this point, we do not completely understand the significance of this specific rate, and why this is weakly L-dependent. 

\section{Conclusion}
\label{conclusion}
In this work, we present a novel model system that allows a continuous crossover from a finite three dimensional liquid to a mean field like system by introducing pseudo neighbours that enhance the effective connectivity without distorting the local structure. Previous studies have shown that pseudo neighbours slow down the dynamics, even though the local structure remains unchanged from the parent KA model \cite{ujjwal_mean_field}. It was also observed that the configurational entropy, calculated via thermodynamic integration, vanished more rapidly than the slowing down of dynamics, resulting in a breakdown of the well known correlation between thermodynamics and dynamics via the Adam Gibbs relation. Interestingly, entropy estimated using the phenomenological 2PT method\cite{ujjwal_thermodynamics}, based on the velocity auto-correlation function, remained higher and continued to correlate with dynamics via the Adam Gibbs relation. This indicates that pseudo neighbours have a weaker impact on local dynamic observables, such as the velocity correlation and self and collective overlap functions. However, like in thermodynamics, we find that the pseudo neighbours also have a much stronger influence on long-time behaviours like the onset of diffusion \cite{ujjwal_mean_field}, which depends on the bond breakage dynamics \cite{smarjit_timescale}. Thus, we probe the connection between thermodynamics and the bond breakage dynamics in the supercooled regime, particularly focusing on the signature of a thermodynamic transition characterised by the vanishing of configurational entropy at $T_{K}$. 

 In the mean field system, two types of bonds are present, the real–real bonds (between actual particles) and real-pseudo bonds (between a real particle and a pseudo neighbour). We find that the real–real bond breakage time correlation function, although slower than the same in the KA system, ultimately saturates at the same long time value. Interestingly, we find that the bond breakage time correlation function in the KA model itself does not fully decay to zero, as also observed in earlier studies, but not specifically reported \cite{Ryoichi_Yamamoto1997,smarjit_timescale}. This indicates the presence of residual, long lived bonds. However, our analysis reveals that these residual real–real bonds, both in the KA and in the mean field systems, are few in number, weakly temperature dependent, and diminish with increasing system size. As a result, they do not significantly impact the dynamics.

In contrast, the real–pseudo bonds exhibit markedly different behaviour. Their bond breakage correlation function saturates at a much higher value, indicating a larger number of residual pseudo bonds. These persistent real–pseudo bonds increase with decreasing temperature, and at low temperatures, they also grow with system size. This persistence of a finite number of bonds prevents the extraction of a conventional relaxation timescale from the bond breakage dynamics but highlights the importance of the number of surviving bonds as a dynamic measure. We further demonstrate that the number of surviving real and pseudo bonds can be calculated via statistical analysis using the information of the structure of the liquid, and these match the values obtained from simulations.

A central contribution of this work is the proposed analytical expression that relates the number of surviving pseudo bonds to the thermodynamic transition temperature $T_{K}$. The expression reveals that $T_{K}$ receives an additive contribution from both real and pseudo-neighbours, the first term corresponds to the baseline value, which is the contribution from the real neighbours and the same as the $T_{K}$ in the KA model, while the second term grows linearly with the number of surviving pseudo bonds. This result clearly demonstrates the independent contribution of the surviving pseudo bonds in the thermodynamic transition. Overall, our findings reveal that surviving pseudo bonds act as a robust dynamic signature of the thermodynamic transition. 

Recently, a randomly bonded system has been proposed in which certain particle pairs are permanently linked. This model has been shown to produce ultrastable glasses \cite{ozawa_bond_nature,ozawa_bond_jstat}. Although it shares some conceptual similarities with random pinning, particularly the reduction of configurational degrees of freedom, it differs in important ways \cite{ozawa_bond_jstat}. Notably, the randomly bonded system preserves translational invariance, whereas random pinning explicitly breaks it. Furthermore, in the randomly bonded system, the self and collective dynamics remain coupled, while in randomly pinned systems, they tend to decouple. As discussed earlier, our model shares several similarities with the randomly pinned system, particularly an early vanishing of configurational entropy at relatively high temperatures and the breakdown of the AG relationship \cite{ujjwal_thermodynamics}. However, unlike the pinned system, our model, similar to the randomly bonded case, is translationally invariant, and as shown here, the self and collective dynamics remain coupled. In our mean field model, although there are no permanent bonds, the long lived persistent bonds with pseudo neighbours may functionally behave like soft, reversible bonds, providing the system with features reminiscent of soft bonded systems
These observations suggest that our system may provide an alternative route to realising ultrastable glassy states. Investigating whether this soft-bonded mean field system also exhibits the hallmarks of ultrastability will be an exciting direction for future research.

\appendix
\numberwithin{equation}{section}
\section{Definition}
\label{overlap}
\subsection{Self and collective overlap}
We compute the self part, $Q_{s}$, and the collective part, $Q_{c}$, of the overlap function, which represents a two-point time correlation of the local density\cite{walter_Tg}. The self part is defined as, 
\begin{equation}
Q_{s}(t) =\frac{1}{N} \Big \langle \sum_{i=1}^{N} \omega (|{\bf{r}}_i(t)-{\bf{r}}_i(0)|)\Big \rangle \quad 
\label{self}
\end{equation}

The function $\omega(x)$ is defined such that it equals 1 when $0\leq x\leq a$, and 0 otherwise. The parameter a, set to 0.3, is slightly larger than the typical cage size, which is inferred from the height of the plateau in the mean square displacement at intermediate times \cite{Kob1995}. Consequently, the function $Q_{s}$ indicates whether a tagged particle remains within its original cage at time $t$, compared to its position at $t=0$. The collective part is defined as follows,
\begin{equation}
Q_{c}(t) =\frac{1}{N} \Big \langle \sum_{i=1}^{N}\sum_{j=1}^{N} \omega (|{\bf{r}}_i(t)-{\bf{r}}_j(0)|) \Big \rangle \quad 
\label{collective}
\end{equation}
\noindent
\subsection{Radial distribution function}
The partial radial distribution function $g_{\alpha\beta}(r)$ is defined as \cite{Hansen_and_McDonald}, 
\begin{equation}
g_{\alpha\beta}(r) =\frac{V}{N_{\alpha}N_{\beta}}\Big<\sum_{i=1}^{N_{\alpha}} \sum_{j=1, j \neq i}^{N_{\beta}} \delta (r- r_i +r_j) \Big>  
 \label{gr_2}
\end{equation}
\noindent

\begin{figure}[h!]
 \centering
 \includegraphics[width=0.5\textwidth]{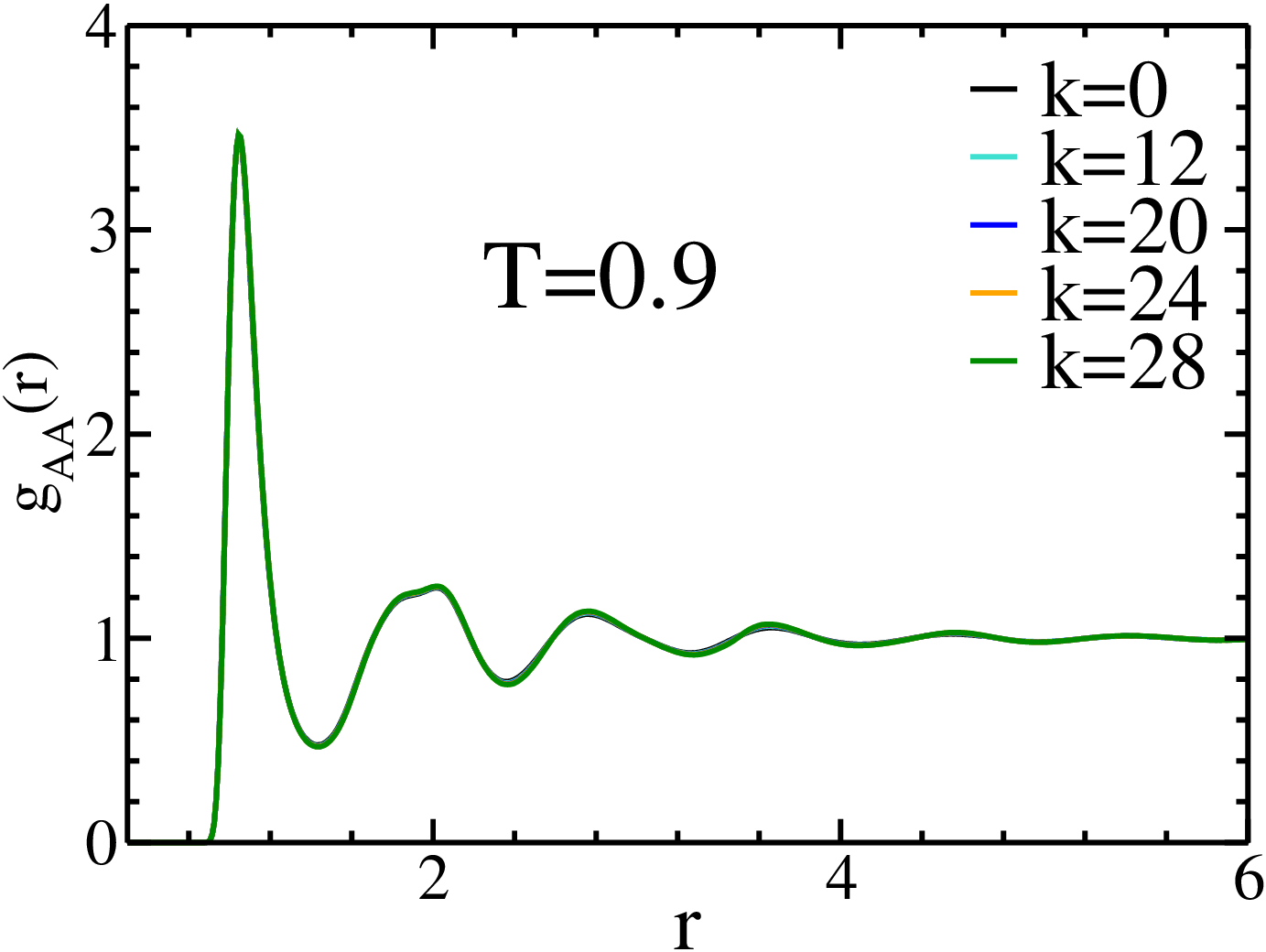}
 \caption{The radial distribution function for a system with $L=2.5$ and pseudo-neighbours $k=0,12,20,24$, and $28$ at $T=0.9$ shows that the structure remains invariant under the introduction of the pseudo-neighbours.}
 \label{rdff}
 \end{figure} 
\noindent 
where $V$ is the system volume, and $N_{\alpha}$ and $N_{\beta}$ represent the number of particles of types $\alpha$ and $\beta$, respectively.
The overall (effective one-component) radial distribution function $g(r)$ for a binary system is given by \cite{Hansen_and_McDonald}:
\begin{equation}
 g(r) = \sum_{\alpha,\beta=1}^{2} \chi_{\alpha} \chi_{\beta} g_{\alpha\beta}(r)
 \label{gr}
\end{equation}
where $\chi_\alpha$ and $\chi_\beta$ are the mole fractions of particle types $\alpha$ and $\beta$, respectively.

\section{Entropy Calculation}
\label{sec_entropy}
In this work, we have used well-known Density Temperature integration methods for the calculation of the total entropy ($S_{tot}$) of the system.
Below, we outline the various quantities necessary for calculating the entropy using this approach.

\subsubsection{Entropy}
The entropy of the system is obtained from the density temperature integration method \cite{sciortino2000thermodynamics,sastry2000onset_TI} In this method, the entropy is initially evaluated at a high temperature and low reduced density for each model, where the system can be assumed to behave as an ideal gas and the ideal gas entropy, $S_{ideal}$, is given by, 
\begin{equation}
S_{ideal}(T,V) = \frac{5}{2}-\ln (\rho) + \frac{3}{2}\ln \Big(\frac{2\pi T}{h^2}\Big) + \frac{1}{N}\ln \frac{N!}{N_{A}!N_{B}!} 
\label{S_ideal_eq}
\end{equation}
\noindent where $N=N_{A}+N_{B}$ is the total number of particles, $V$ is the volume of the system and $h$ is the Planck constant. $N_{A}$ and $N_{B}$ are number of particles of type A and B. The last term contributes to the mixing entropy.

Entropies at any other state points, relative to this reference ideal state point, can be calculated using a combination of isochoric
and isothermal paths, ensuring that no phase transitions occur along the path \cite{chakravarty_DTI_method}. The entropy change along an isothermal path is given by,
\begin{equation}
 S(T^\prime,V)- S(T^\prime,V^\prime)=\frac{U(T^\prime,V)- U(T^\prime,V^\prime)}{T^\prime}+\int_{V^\prime}^{V}\frac{P(V)}{T^\prime}dV,
\label{isothermal}
\end{equation}
Along the isochoric path, it is given by,

\begin{equation}
 S(T,V)- S(T^\prime,V)=\int_{T^\prime}^{T}\frac{1}{T} \left(\frac{\partial U}{\partial T}\right)_V dT.
 \label{isochoric}
\end{equation}
\noindent

In the supercooled liquid phase, the configurational space is partitioned into inherent structure minima and the vibrational motions surrounding them \cite{sciortino_prl_1999,sastry_inherent_1998}. The configurational entropy, $S_{c}(T,V)$, of the system is defined as the logarithm of the number of these inherent structure minima. It can be determined by subtracting the vibrational entropy,$S_{vib}(T,V)$,
from the total entropy,$S(T,V)$:

\begin{equation}
\begin{aligned}
 S_{c}(T,V) = S(T,V) - S_{vib}(T,V) \\
\end{aligned}
\label{S_c_eq}
\end{equation}
\noindent

The vibrational entropy is calculated by making a harmonic approximation about a local minimum \cite{sastry2000liquid,Sastry2001,sengupta2011dependence}.  To obtain the vibrational frequencies, we calculate the Hessian and then diagonalise it. Once we obtain the vibrational frequencies, $S_{vib}$ is calculated using the following equation,

\begin{equation}
S_{vib}(T,V) = \frac{3}{2}\ln \Big(\frac{2\pi T}{h^{2}} \Big)+ \frac{\ln(V)}{N}+\frac{1}{2N}\sum_{i=1}^{3N-3}\ln \Big(\frac{2\pi T}{{\omega_{i}}^{2}} \Big)-\frac{3}{2N} + 3
\label{S_vib_eq}
\end{equation}
\noindent

\begin{figure}[h!]
 \centering
 \includegraphics[width=0.5\textwidth]{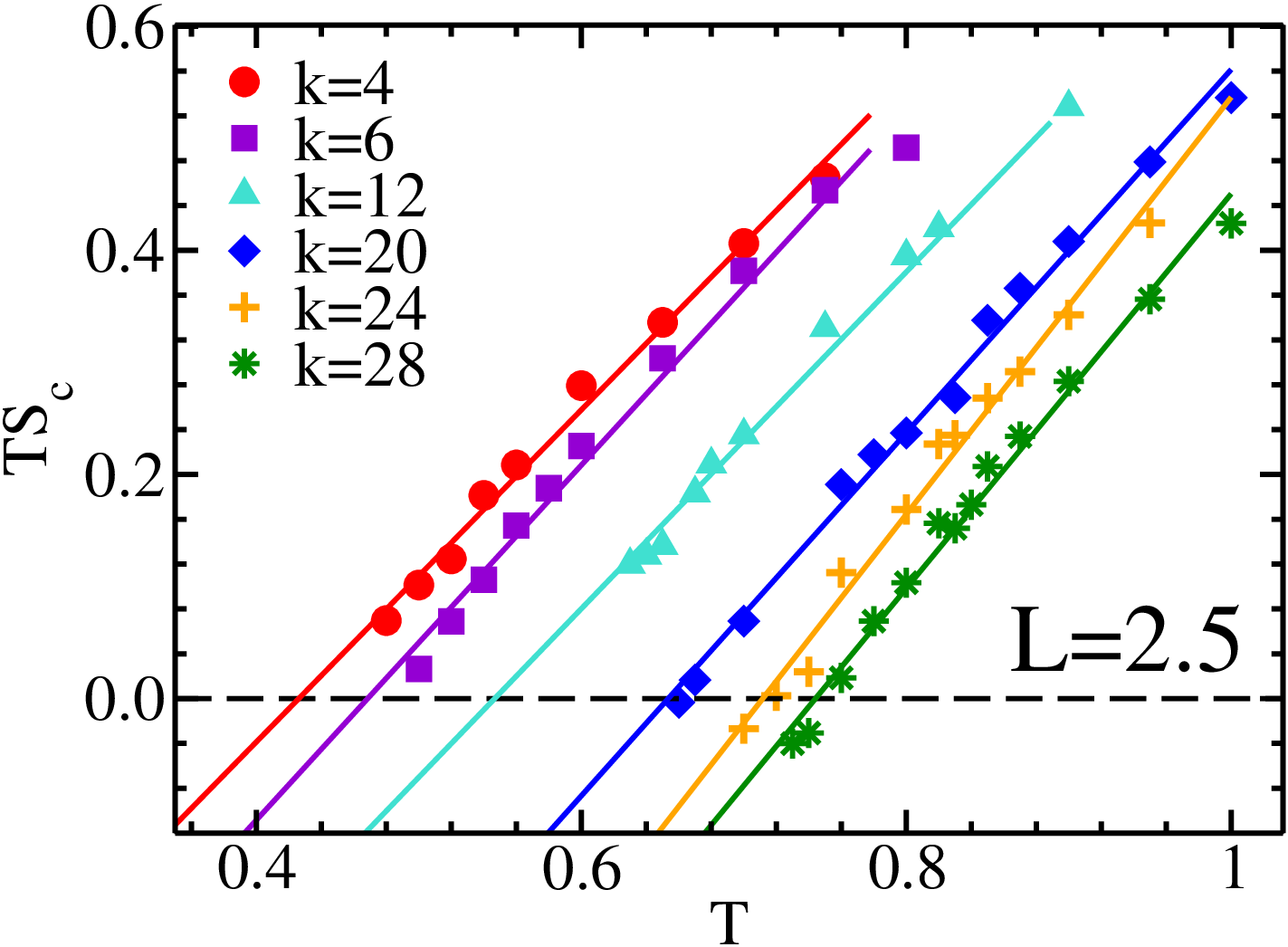}
 \caption{The temperature dependence of the configurational entropy $S_{c}$ in the mean field system. $TS_c$ vs $T$ for $L=2.5$ and $k=4,6,12,20,24$, and $28$. $S_c$ is calculated using the Density Temperature integration method. The value of $T_K$ increases with increasing k.}
 \label{config_ent_2.5}
 \end{figure}

\begin{figure}[h!]
 \centering
\vspace{0.5cm}
 \includegraphics[width=0.5\textwidth]{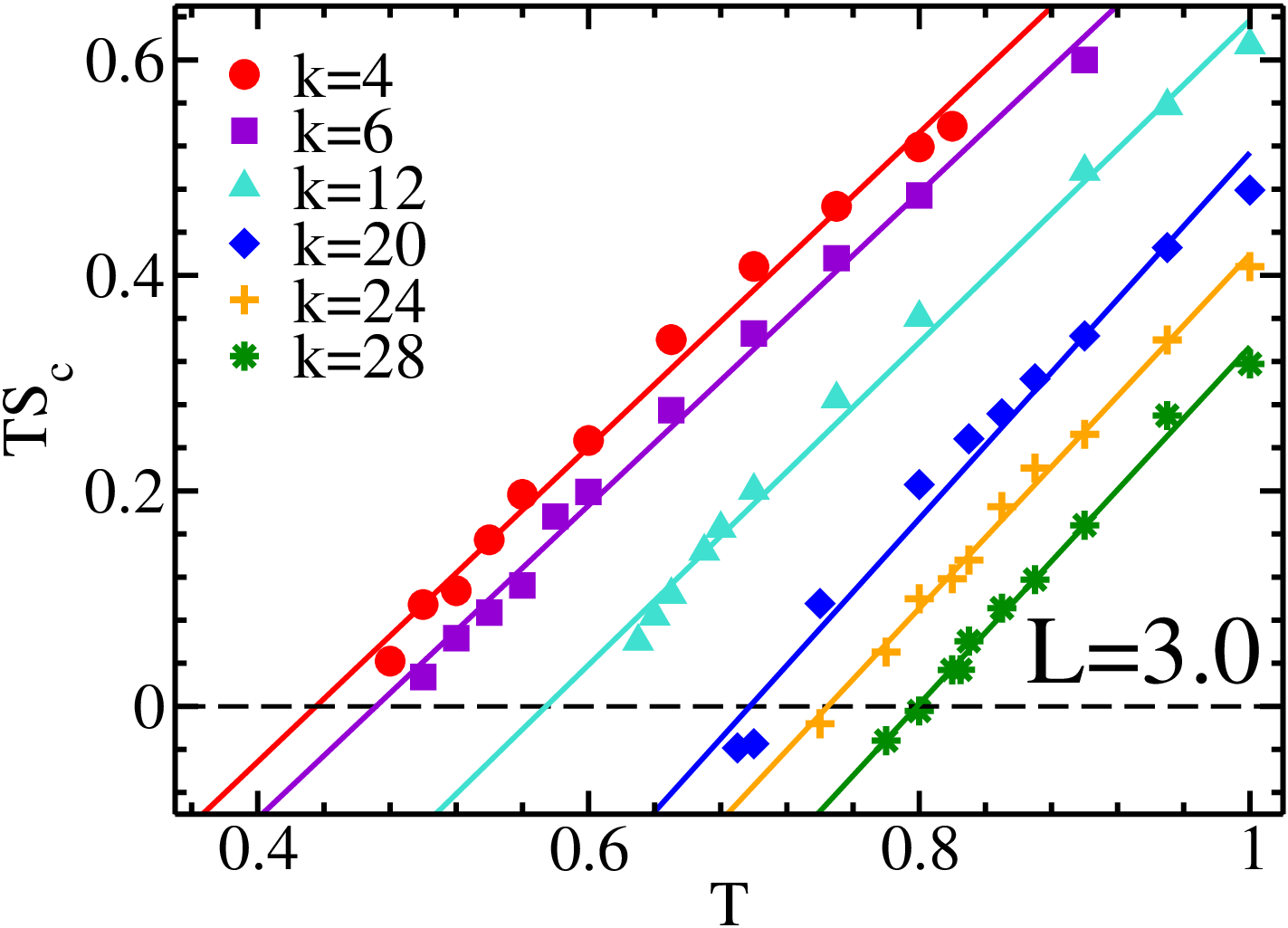}
 \caption{The temperature dependence of the configurational entropy $S_{c}$ in the mean field system. $TS_c$ vs $T$ for $L=3$ and $k=4,6,12,20,24$, and $28$. $S_c$ is calculated using the Density Temperature integration method. The value of $T_K$ increases with increasing k.}
 \label{config_ent_3.0}
 \end{figure}

The $TS_{c}$ vs T plot which allows us to find the $T_{K}$ values (given in Table (\ref{Tk_table})) are given in Fig.\ref{config_ent_2.5} and Fig.\ref{config_ent_3.0}.

\vspace{8mm}

{\bf ACKNOWLEDGMENT}\\
 S.~M.~B. Thanks, Science and Engineering Research Board (SERB, Grant No. SPF/2021/000112 ) for the funding. E.~A. Thanks, SERB and CSIR-NCL, for fellowships. U.K.N. acknowledges Japanese Government Monbukagakusho Scholarship (MEXT), JST CREST Grant Number JPMJCR2095, Kyoto University. S.K. thanks DST-INSPIRE for the fellowship. S.~M.~B. would like to thank Walter Kob and Gill Tarjus for discussions.  \\[4mm]

{\bf AVAILABILITY OF DATA}\\
The data that supports the findings of this study is available from the corresponding author upon reasonable request.\\[3mm]

\textbf{References}

%\bibliographystyle{h-physrev}
%\nocite{*}
%\bibliography{Reference}

\end{document}